\documentclass{aa}

\usepackage{graphics}

\begin{document}
\thesaurus{02(12.03.4; 12.07.1; 12.04.1; 11.08.1)}
\title{Lensing effects in an inhomogeneous universe}

\author{Lars Bergstr{\"o}m 
  \and Martin Goliath 
  \and Ariel Goobar 
  \and Edvard M\"ortsell \thanks{\emph{Send off-print requests to}:
    Edvard M\"ortsell \newline (edvard@physto.se)}
}
\institute{Department of Physics, Stockholm University, \\
  Box 6730, S--113 85 Stockholm, Sweden
}
\date{Received / Accepted}

\maketitle

\begin{abstract}
  Recently, Holz \& Wald have presented a new method for determining
  gravitational lensing effects on, e.g., supernova luminosity versus
  redshift measurements in inhomogeneous universes. In this 
  paper, their method is generalized in several ways: First, the
  matter content is allowed to consist of several different types of
  fluids, possibly with non-vanishing pressure. Second, besides
  lensing by simple point masses and singular isothermal spheres, the
  more realistic halo dark matter distribution proposed by Navarro, Frenk
  \& White (NFW), based on N-body simulation results, is treated.  
  We discuss various aspects of the
  accuracy of the method, such as luminosity corrections, and
  statistics, for multiple images. 
  We find in agreement with other recent work that a large 
  sample of supernovae at large redshift could be used to extract
  gross features of the mass distribution of the lensing dark
  matter halos, such as the existence of a large number of 
  point-like objects. The results for the isothermal sphere and
  the NFW model are, however, very similar if normalized to the
  observed luminosity distribution of galaxies. We give convenient
  analytical fitting formulas for our computed lensing probabilites
  as a function of magnification, for several redshifts. \\[\baselineskip]
\end{abstract}

\section{Introduction}

Ever since the foundation of modern cosmology based on general
relativity, there have been attempts to determine the cosmological
parameters of our universe. Gravitational lensing has been
recognized as one of several such tools. For example,
the results of radio surveys for gravitational lenses
have recently been used to place significant
constraints on cosmological models (Falco et al. \cite{radiolens}).
It has also been realised  
that observations of supernovae at high redshift can be
used for this purpose (Goobar \& Perlmutter \cite{art:GoobarPerlmutter1995}),
in particular for determining the value of the cosmological
constant. In fact, several collaborations with this in mind are in
progress, and the first sets of data show an intriguing hint of a
non-vanishing cosmological constant in a universe consistent with having
a flat geometry (see, e.g., Riess et al. \cite{art:Riess-et-al1998};
Perlmutter et al. \cite{art:Perlmutter-et-al1999}). 
Although the two groups which have published results mutually agree 
on the best-fit parameters, it is important to note that the effects
of geometry are small (on the order of half a magnitude), and the need
to go to even higher redshift to get larger effects is obvious.
When observing such distant sources, at redshift greater than 
unity, it is necessary to estimate the effects of lensing due to
inhomogeneities in the matter distribution. This will be of growing
importance when supernovae at very high redshifts become accessible,
e.g., with NGST, the Next Generation Space Telescope
(Miralda-Escud\'e \& Rees \cite{MI97}; Dahl\'en \& Fransson
\cite{DA99}) or with the dedicated supernova search satellite, SNAP
(Perlmutter et al., private communication).

The literature on gravitational lensing is quite rich. Much of
the history and developments up to the early  1990s can be found in
the excellent textbook by Schneider, Ehlers \& Falco
(\cite{book:Schneider}). A new method for examining lensing effects
has recently been proposed by Holz \& Wald (\cite{art:HolzWald1998}; HW),
one which has the virtue of lending itself easily to numerical calculations. 
In that work it is also shown that given a large and deep enough
sample of standard candle supernovae, one may in principle distinguish
between different matter distributions in the galactic halos responsible
for the lensing. In particular the case of point masses gives a quite
different distribution of magnifications (or de-magnifications) 
compared to the case of singular isothermal spheres (see
also Metcalf \& Silk \cite{metcalf}).

The method of HW can be summarized as follows: First, a
Friedmann-Lema\^{\i}tre (FL) background geometry is
selected. Inhomogeneities are accounted for by specifying a matter
distribution in cells which have an average energy density equal to
that of the underlying FL model. Following the history of observed
light from a distant source, a light ray is traced backwards to the
desired redshift by being sent through a series of cells, each time
with a randomly selected impact parameter with respect to the
matter distribution in the cell. Between each cell, the
FL background is used to update the scale factor
and the expansion rate. By using Monte Carlo techniques to trace a
large number of such light rays, statistics for the apparent
luminosity of an ensemble of sources at a given redshift is obtained. 

The purpose of this paper is to generalize the method of HW in a
number of ways and to investigate some issues related to the different 
lensing signatures in the luminosity distribution of supernovae
caused by different mass distributions of the intervening lens 
population.
The outline is as follows. In Sect.~\ref{sec:cosmodel}
it is demonstrated how the HW method can be generalized from matter in
dust form to perfect fluids with pressure. 
Sect.~\ref{sec:area} is concerned with the 
interpretation of results as probability distributions.
Sect.~\ref{sec:devi}
presents the geodesic deviation and in 
Sect.~\ref{sec:HWsummary}, we summarize and discuss some of the 
conceptual issues in the analysis of HW. 
In Sect.~\ref{sec:NFW}, we include the possibility of using
the more realistic halo matter distribution proposed 
by Navarro et al. (\cite{art:NFW}; NFW) based on their N-body
simulation results, and in Sect.~\ref{sec:massdist} we
discuss the mass distribution of lensing galaxies. By using realistic
density profiles and mass distributions of dark matter halos, we
can obtain high accuracy results on gravitational lensing  without the need
of using the extensive full data sets of N-body simulations, as in
some other methods (Wambsganss et al. \cite{cen};
Premadi et al. \cite{premadi}; Jain et al. \cite{jain}). 
One strength of our ray-tracing
Monte Carlo method is further that it can be continuously refined as
more observational information, e.g., on galaxy distributions is
obtained. Also, effects such as absorption by dust and other
possibly $z$-dependent effects can be straightforwardly
added to the algorithm.
In Sect.~\ref{sec:results}, some results from the generalizations of the 
method of HW are presented as luminosity distributions.
In Sect.~\ref{sec:limit},
we investigate some limitations of the numerical implementation 
of the method and
in Sect.~\ref{sec:mi}, we discuss multiple images. 
Sect.~\ref{sec:cc} investigates different consistency checks,
Sect.~\ref{sec:formula} contains convenient analytical
fits to our numerical results and
the paper is concluded with a discussion in Sect.~\ref{sec:disc}.

\section{Cosmological model}\label{sec:cosmodel}

The starting point of HW is a Newtonianly perturbed FL universe. The
line element of Robertson-Walker form is then
\begin{equation}\label{eq:line-element}
  ds^2=-(1+2\phi)dt^2 +(1-2\phi)a^2(t)dl^2_k ,
\end{equation}
where $a$ is the scale factor, and $dl^2_k$ is a constant-curvature
three-geometry, for which $k\in\{0,\pm1\}$ indicates the sign of the
curvature. (We use geometrized units such that $G=c=1$).
The perturbation $\phi$ is assumed to satisfy a number of
properties that 
will be employed when determining dominant terms. Firstly,
$|\phi|\ll1$, reflecting that $\phi$ indeed is a perturbation. Secondly,
inhomogeneities are assumed to be significant, so that $\partial
\phi/\partial t$ is small compared to spatial derivatives. Thirdly,
second-order spatial derivatives will be assumed to dominate over
products of first-order ones 
(e.g., $(\partial_x\phi)^2\ll |\partial_{xx}\phi|$).
In this paper, we have generalized the treatment of HW so that 
the matter content of the model can be taken to be a number of
perfect fluids: 
\begin{equation}
  T_{ab}=\sum_i \rho_i u_au_b + p_i(u_a u_b + g_{ab}) ,
\end{equation}
where $T_{ab}$ is the stress-energy tensor, $\rho_i$ is the energy 
density, $p_i$  the pressure of
fluid $i$, $u^a$ is the (normalized) fluid four-velocity, and $g_{ab}$ is
the metric corresponding to the line element (\ref{eq:line-element}).
Cosmologically reasonable equations of state
are linear barotropic ones: $p_i=\alpha_i\rho_i$.
In this paper, the $\alpha_i$'s are constants with $-1\leq\alpha_i\leq0$. 
However, it is straightforward to allow for the equations of state to
evolve with redshift, 
i.e., $\alpha_i =\alpha_i (z)$.
 
HW considered dust models with a cosmological constant, which corresponds
to a two-fluid model, the first fluid having $\alpha_1=0$, and the
second fluid homogeneously distributed with $\alpha_2=-1$ and
$\Lambda=8\pi\rho_2$.
The field equations, corresponding to Eqs.~(HW7,8), become\\ 
{\it The Friedmann equation:}
\begin{equation}\label{eq:fe}
  3\left(\frac{\dot{a}}{a}\right)^2=3H^2=8\pi\sum_i\rho_i-
  \frac{3k}{a^2}-2\nabla^2\phi ,
\end{equation}
{\it The Raychaudhuri equation:}
\begin{equation}\label{eq:re}
  3\frac{\ddot{a}}{a}=-3qH^2=-4\pi\sum_i(\rho_i+3p_i)+\nabla^2\phi ,
\end{equation}
where the definitions of the Hubble scalar $H$ and
deceleration parameter\footnote{The name is of historical origin. Actually,
according to the current supernova cosmology data the expansion of
the universe is presently accelerating, i.e. $q<0$.}
 $q$ have been made explicit, and where
$\nabla^2$ denotes the spatial projection of contracted covariant
derivatives. The pressure does not enter the Friedmann equation, i.e.,
the Hubble scalar is entirely given by the energy densities and the
curvature. The deceleration, on the other hand, is affected by the
presence of pressure. Comparing the above field equations with the
field equations for the corresponding FL
background (with averaged densities and pressures $\bar{\rho}_i$, 
$\bar{p}_i$) it can be shown [Eq.~(HW11)] that $\phi$ satisfies a
Poisson equation $\nabla^2\phi=4\pi\sum_i\delta\rho_i$, where
$\delta\rho_i=\rho_i-\bar{\rho}_i$, while $\delta p_i=p_i-\bar{p}_i$
vanishes. Thus, the pressures result in homogeneous contributions, and
the specification of an equation of state only makes sense in terms of
the averaged energy density: $p_i=\bar{p}_i=\alpha_i\bar{\rho}_i$. 

From the field equations for the background model follows\\ 
{\it The energy conservation equation:}
\begin{equation}\label{eq:ec}
  \sum_i\dot{\bar{\rho}}_i + 3(\bar{\rho}_i+\bar{p}_i)H=0 .
\end{equation}
If the fluids are assumed to couple only to gravity, they will be
separately conserved, and each component in the sum of
Eq.~(\ref{eq:ec}) vanishes. Note that it is only for dust ($p=0$)
that the mass within a volume co-moving with the Hubble flow is
constant ($\bar{\rho}\propto a^{-3}$). For other types of matter, the mass
varies with time because of the work done by the pressure. Also note
that $p=-\rho$ results in a constant energy density. Thus, a
homogeneously distributed fluid with $p=-\rho$ is equivalent
to a cosmological constant, or in terms of quantum field models, to
vacuum energy. It is convenient to introduce density
parameters 
\begin{equation}
\Omega_i=8\pi\bar{\rho}_i/3H^2,
\end{equation}
 which depend on time and
therefore on redshift. Assuming the above discussed
equations of state  the time dependence is governed by 
the following set of conservation
equations (see, e.g., Goliath \& Ellis \cite{art:GoliathEllis1999})
\begin{equation}
  \dot{\Omega}_i=\left[2q-(3\alpha_i+1)\right]\Omega_iH .
\end{equation}

HW proceed by performing a number of coordinate transformations to
co-moving coordinates on isotropic form, Eq.~(HW16). The inclusion of
pressure results in a modification of this line element:
\begin{eqnarray}\label{eq:lineco}
  ds^2=&-&\left[1+2\Phi+4\pi R^2\sum_i p_i\right]
  dT^2 \nonumber \\
  &+&\left[1-2\Phi\right](dX^2+dY^2+dZ^2) ,
\end{eqnarray}
where $R^2\approx X^2+Y^2+Z^2$, and the effective potential
\begin{equation}
\Phi=\phi+\frac{2\pi R^2}{3}\sum_i\rho_i
\end{equation} satisfies the Poisson
equation 
\begin{equation}\nabla^2\Phi=4\pi\sum_i\rho_i. 
\end{equation}


The lookback time of an event at redshift $z_1$ is the time difference
(with respect to the coordinate $t$) between the event and the
present. By using $1+z=a_0/a$, we get a relation between the lookback
time and the redshift for a FL universe: 
\begin{equation}
  \Delta t=\int_{a_1}^{a_0}\frac{da}{\dot{a}}
  =\int_0^{z_1}\frac{dz}{(1+z)H(z)} .
\end{equation}
To obtain $H(z)$, the Friedmann equation, Eq.~(\ref{eq:fe}),
is expressed in terms of redshift and the present densities
$\rho_{0i}$. To do so, first note that the energy conservation
equation, Eq.~(\ref{eq:ec}), together with the equation of state
gives
\begin{equation}
  \bar{\rho}_i=\bar{\rho}_{0i}\left(\frac{a}{a_0}\right)^{-3(1+\alpha_i)}
  =\bar{\rho}_{0i}(1+z)^{3(1+\alpha_i)} .
\end{equation}
The Hubble scalar can then be expressed
\begin{equation}
  H(z)=H_0\sqrt{\sum_i\Omega_{0i}(1+z)^{3(1+\alpha_i)}+
    \Omega_{0k}(1+z)^2} ,
\end{equation}
where we introduced the curvature density parameter
\begin{equation}
  \Omega_{0k}=1-\sum_i\Omega_{0i}=-\frac{k}{H_0^2a_0^2} .\label{eq:ook}
\end{equation}
The luminosity distance to the event at redshift $z_1$ is given by a
similar expression (see, e.g., Bergstr\"om \& Goobar
\cite{book:BergstromGoobar1999})
\begin{eqnarray}
  d_L&=&a_0(1+z)\ \mbox{\Large {\it f}} \,\left(\frac{1}{a_0}
  \int_0^{z_1}\frac{dz}{H(z)}\right) , \\
  f(x)&=&\left\{
  \begin{array}{ll}
    \sin(x) , & k=1 \\
    x , & k=0 , \\
    \sinh(x) , \quad & k=-1 \\
  \end{array}
  \right.
\end{eqnarray}
Note that, when $k=\pm1$, the present scale factor can be expressed 
[using (\ref{eq:ook})] as
$a_0=(H_0\sqrt{|\Omega_{0k}|})^{-1}$.

\section{Area vs. magnification and statistical weighting}\label{sec:area}

\begin{figure}[htp]
  \begin{center}
    \resizebox{\hsize}{!}{\includegraphics{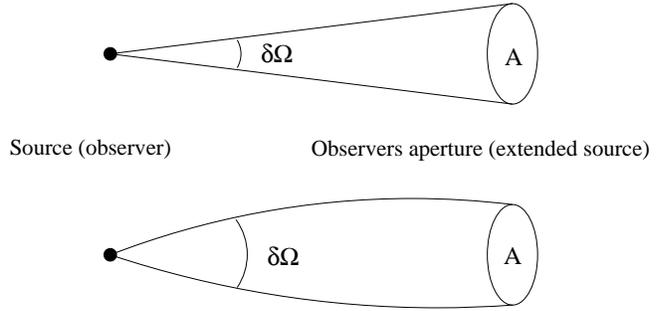}}
    \caption{Geometry of ray bundle.}
    \label{fig:focusing}
  \end{center}
\end{figure}

Consider a narrow (strictly speaking; infinitesimal) ray bundle
connecting a source and an observer 
as depicted in Fig.~\ref{fig:focusing}. The ray bundle can be focused
either at the source or the observer.
The relevant quantity when determining 
luminosity distances is the ratio of the area of a beam and the solid 
angle it subtends. Of course, the smaller the distance, the brighter
the source. 

In the method of HW, a 
beam focused at the observer with infinitesimal $\delta\Omega$ is 
integrated backwards using the geodesic deviation equation, and the 
resulting area at the source sphere is considered. 
In the lower panel of Fig.~\ref{fig:focusing}, the beam has 
undergone additional focusing.
A finite aperture at the observer will detect light from a larger 
$\delta\Omega$ of an isotropically emitting 
source, 
and a finite source will subtend a larger solid angle on the sky.
Since surface 
brightness is conserved, the source will be magnified. The 
magnification, $\mu$, is defined as the ratio of the {\em observed}
flux and the flux of the corresponding image in absence of any
additional focusing. 
\begin{equation}
    \mu =\frac{\delta\Omega}{\delta\Omega_{0}}
    \label{eq:mu}
\end{equation}
where $\delta\Omega_{0}$ denotes the solid angle subtended by the source in 
absence of lenses. 
Hence the
magnification will be proportional to the inverse of the area of the
beam. 
In absence of any lenses, the ray bundle will have an area
corresponding to the so called \emph{empty-beam} area
(see Sect.~\ref{sec:devi}), and we will therefore define the
magnification as  
\begin{equation}
    \mu =\frac{A_{\rm empty}}{A}.
    \label{eq:mu2}
\end{equation}
For further reference, see Schneider, Ehlers \& Falco
(\cite{book:Schneider}).

\begin{figure}[htp]
  \begin{center}
    \resizebox{\hsize}{!}{\includegraphics{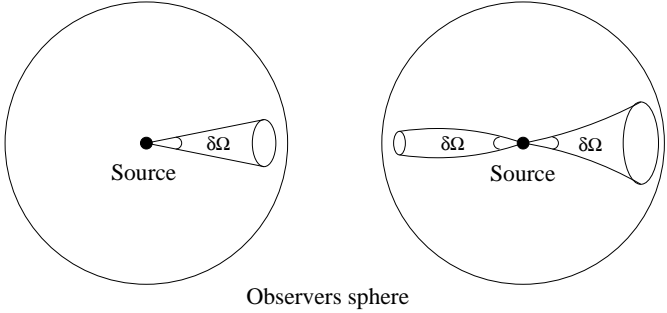}}
    \caption{Different ray bundles with constant $\delta\Omega$}
    \label{fig:focusing3}
  \end{center}
\end{figure}

As noted in HW, some care is needed when interpreting 
results as probability distributions. This is due to the fact that 
individual ray bundles do not correspond to random source positions.
This is illustrated in the left panel of Fig.~\ref{fig:focusing3} where 
we consider the situation of a ray bundle originating from an observer
at $z=0$ at the center of a source sphere at some fixed redshift. 

From the right panel of Fig.~\ref{fig:focusing3} we see that focused beams 
sample a smaller fraction of the source sphere. 

Since the fraction of the 
source sphere being sampled by a beam is proportional to the area of the 
beam, i.e., inversely proportional to the magnification,
magnified sources will be overrepresented if not
compensated for. In HW, this is done by convoluting the obtained
probability distribution, $n_{0}(A)$, with a factor proportional to
$A$, thus obtaining a new distribution $n(A)\propto A\, n_{0}(A)$.
The method we use is very similar in that we designate a probability,
proportional to the area of the ray bundle, to keep every event. The
advantage of this rejection method is that each selected event
represents a statistically ``real'' event. 


Further discussions of the statistical weight given to random 
line-of-sights can be found in Ehlers \& Schneider (\cite{EH86}). 

\section{Geodesic deviation}\label{sec:devi}

The geodesic deviation equation for a light ray is (see, e.g.,
Wald \cite{book:Wald1984}) 
\begin{eqnarray}
  \frac{d^2\xi^a}{d\lambda^2}&=&-\Delta^a\!_b  \xi^b , \\
  \Delta_{ab}&=&R_{cadb} k^c k^d ,
\end{eqnarray}
where $\xi^a$ is the deviation vector and $k^a$ is a null vector tangent 
to the light ray. The affine parameter $\lambda$ is a
parametrization along the light ray given by $\omega d\lambda =dt$,
where $\omega$ is the frequency of the photons
in the beam.  
It is instructive to calculate the
deviation in an exact (homogeneous) FL universe, analogous to
Eq.~(HW21):
\begin{equation}\label{eq:FLdev}
  \frac{d^2\xi^\alpha}{d\lambda^2}=-4\pi\omega^2\sum_i(\rho_i+p_i)
  \xi^\alpha . 
\end{equation}
Thus, the pressure will cause additional focusing of the
light beam. Note that a cosmological constant, which corresponds to
$p=-\rho$, does not affect the deviation.

To derive the deviation experienced when propagating through a cell in
the inhomogeneous model, HW start by writing the deviation equation in
terms of a $2\times 2$ matrix ${\cal A}^\alpha\!_\beta$, such that
\begin{equation}\label{eq:deviA}
  \frac{d^2{\cal A}^\alpha\!_\beta}{d\lambda^2}=
  -\Delta^\alpha\!_\gamma {\cal A}^\gamma\!_\beta ,
\end{equation}
where $\alpha,\beta\in\{X,Y\}$ (HW28). The properties of the geodesic
deviation are contained in 
${\cal A}^\alpha\!_\beta$.\footnote{This matrix is
conventionally discussed in connection with conjugate points
for null geodesics, (see, e.g., Wald \cite{book:Wald1984};
Hawking \& Ellis \cite{book:HawkingEllis1973}).}
For example, the area (with orientation) of the beam is given by
$\det {\cal A}$. 
In the case of the homogeneous background Friedmann-Lema\^{\i}tre cosmology,
we denote the corresponding reference area $A_{\rm FL}$, which will
of course depend on the redshift, but also on the cosmological
parameters of the FL background model. Also, the area of a light beam
that propagates without encountering any matter will be denoted
$A_{\rm empty}$. Sometimes $A_{\rm FL}$ and $A_{\rm empty}$ are
referred to as the \emph{filled-beam} and \emph{empty-beam} areas,
respectively.

The deviation equation, Eq.~(\ref{eq:deviA}), can now
be written as a set of difference equations (HW35,36)
\begin{eqnarray}
  \label{eq:AdA}
  ({\cal A}^\alpha\!_\beta)_1&=
  ({\cal A}^\alpha\!_\beta)_0 +
  \Delta\lambda(d{\cal A}^\alpha\!_\beta/d\lambda)_0 , \\
  (d{\cal A}^\alpha\!_\beta/d\lambda)_1&=
  (d{\cal A}^\alpha\!_\beta/d\lambda)_0 -
  \omega J^\alpha\!_\gamma({\cal A}^\gamma\!_\beta)_0 ,\label{eq:HW36}
\end{eqnarray}
with
\begin{equation}
  J^\alpha\!_\beta=\frac{1}{\omega} \int_c d\lambda\Delta^\alpha\!_\beta ,
\label{eq:jab}
\end{equation}
where the integration is along a straight line through the cell.
Next, $\Delta_{\alpha\beta}$ is calculated from the line element
Eq.~(\ref{eq:lineco}) 
\begin{equation}
  \Delta_{\alpha\beta}\!=\!\omega^2\!
  \left\{\!2\partial_\alpha\partial_\beta\Phi\!+\!
    \left[Z^a\partial_a(Z^b\partial_b\Phi)\!+\!4\pi\!\sum_i p_i\right]
      \!\eta_{\alpha\beta}\!\right\} ,
\end{equation}
where $\eta_{ab}$ is the Minkowski metric associated with the
coordinates $(T,X,Y,Z)$. Thus, with non-vanishing pressure, there will
be an additional contribution in $J^\alpha\!_\beta$ that takes the form 
\begin{equation}
  J^\alpha_p\!_\beta=
\left(\sum_i J_{p_i}\right)\,\delta^\alpha\!_\beta ,
\end{equation}
with
\begin{eqnarray}\label{eq:Jp}
  J_{p_i}&=&4\pi\omega p_i \int_c d\lambda \nonumber \\
  &=& 4\pi p_i \int_c dZ \nonumber \\
  &=& 4\pi p_i \, 2R_c\sqrt{1-\frac{b^2}{R_c^2}} \nonumber \\
  &=& \frac{6 p_i V_c}{R_c^2}\sqrt{1-\frac{b^2}{R_c^2}} ,
\end{eqnarray}
where $b$ is an impact parameter, and $R_c$ and $V_c$ are the radius and
volume of the cell, respectively. Assuming linear barotropic equations
of state, $p_i=\alpha_i\bar{\rho}_i$, Eq.~(\ref{eq:Jp}) becomes
\begin{equation}
  J_{p_i}=\alpha_i \frac{6 M_i}{R_c^2}\sqrt{1-\frac{b^2}{R_c^2}} ,
\end{equation}
where $M_i$ is the mass of matter component $i$ in the cell.

Let us now examine the above result for the filled-beam case.
Eq.~(HW39) gives the quantity
$J^\alpha\!_\beta$ for matter uniformly distributed in a ball of
radius $d$ at the center of a cell of radius $R_c$. Setting $d=R_c$ in
this equation, and adding our derived pressure contribution
$\sum_i J_{p_i}\delta^\alpha\!_\beta$ to this yields
\begin{eqnarray}
  (J^{\rm FL})^X\!_X &=&(J^{\rm FL})^Y\!_Y=\sum_i
  (1+\alpha_i)\frac{6 M_i}{R_c^2}\sqrt{1-\frac{b^2}{R_c^2}} , \nonumber \\
  (J^{\rm FL})^X\!_Y &=& (J^{\rm FL})^Y\!_X = 0 ,
\end{eqnarray}
from which it is seen that a cosmological constant ($\alpha=-1$) does
not contribute, in consistency with Eq.~(\ref{eq:FLdev}).

As an interesting example, consider a flat cosmological model in
which ordinary matter ($\alpha=0$) constitutes only 10 \% of the total
matter content, i.e., $\Omega_M=0.1$, while the rest is made up of
some homogeneously distributed ``X-matter'' with $\Omega_X=0.9$. The ordinary
matter is assumed to be in the form of point masses (we will return later
to more realistic mass distributions). For
this cosmology, Garnavich et al. (\cite{Garnavich}) have put an upper
limit on the parameter of the 
equation of state for an X-matter component using Type Ia supernova data.
Their limit is $\alpha_X <-0.60$ at 95 \% confidence level. Similarly, the
Supernova Cosmology Project
(Perlmutter et al. \cite{art:Perlmutter-et-al1999})  
has deduced a limit of $\alpha_X <-0.44$ (95 \% cl)
for any value of $\Omega_M$.  

\begin{figure}
  \resizebox{\hsize}{!}{\includegraphics{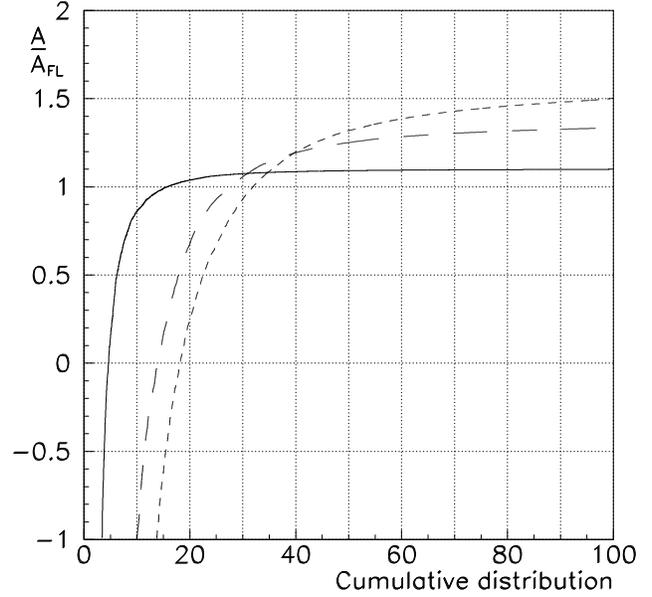}}
  \caption{Beam areas normalized to the filled-beam value of the
    underlying Friedmann-Lema\^{\i}tre model $A_{\rm FL}$ for 10\,000 
    sources at redshift $z=3$ in a $\Omega_M=0.1$, $\Omega_X=0.9$
    universe. The matter is assumed to be in the form of point masses,
    and results for the equation of state coefficient $\alpha_X$ of the 
    homogeneous $X$ component having the value $0$ (solid line), $-2/3$ 
    (long-dashed line) and $-1$ (short-dashed line) are shown. 
    The horizontal scale gives cumulative percentage
    of beams having an area less than a given value. 
    Negative areas correspond to secondary
    images.}\label{fig:Xmatter} 
\end{figure}

In Fig.~\ref{fig:Xmatter}, we compare the cases of $\alpha_X =-1$, 
corresponding to a cosmological constant, $\alpha_X =-2/3$, 
which may either arise from ``quintessence'' (i.e., a slowly evolving
scalar field) (Wang et al. \cite{steinhardt}) or from a domain wall network
(Battye et al. \cite{spergel}). For reference, we also show the result 
for $\alpha_X =0$, corresponding to dust (i.e., equivalent to an
Einstein--de Sitter model with $\Omega_M=1$, of which 90 \% is
homogeneously distributed). 
The choice of $\Omega_M =0.1$ is made to allow for a direct comparison of
the curve corresponding to $\alpha_X =-1$ in 
Fig.~\ref{fig:Xmatter} with Fig.~7 in HW, with which it agrees well.

Note that the normalization differs between the three cases since
$A_{\rm FL}$ depends on the cosmology. Nevertheless, from the slope of the
curves, one can conclude that lensing effects become more prominent for
large negative $\alpha_X$. This is {\em not} what one would expect from 
studying the geodesic deviation equation, Eq. (\ref{eq:FLdev}), since 
negative pressure causes less focusing of a beam. However, for negative
$\alpha_X$, distances will get larger for a fixed redshift, which means a
larger number of lenses between the source and the observer
(see, e.g., Turner \cite{TU90}).   

\section{Summary of the properties of the lensing model}\label{sec:HWsummary}

The method of Holz \& Wald (\cite{art:HolzWald1998}) 
which we adopt has a number of properties,
which we summarize and comment upon here:

\begin{itemize}
  \item {\bf Local nature of lensing}\\
  If $\phi$ is small within distances much smaller than the Hubble
  radius $R_H=H^{-1}$, then the curvature (and thereby the lensing
  effects) is determined locally within $R_H$. Likewise, if we assume that
  there is no strong correlation on scales greater than a clustering scale
  $R$, it follows that the curvature is determined locally within $R$.
  \item {\bf Relevant scales}\\
  The question now is what relevant clustering scale to use. HW find
  it convenient to employ the typical separation between galaxies, and
  argue that galaxy clustering should have a negligible effect on
  magnification, as shown in some of their simulations. 
 They also argue that the smallest relevant scale for
  matter clumping is of the order $10^{-3}M_\odot$, since objects smaller
   than that would not give a coherent lensing effect over the angular size
of most of the interesting sources.
  \item {\bf Generality}\\
  For the case of point masses, HW demonstrate that their results are
  fairly insensitive to the size of individual point masses. 
  Furthermore, their results
  are rather insensitive to clustering. This is tested by cutting out
  a tube in a uniform matter distribution and distributing ``stars''
  within this tube.
  \item {\bf Largest effects}\\
  Based upon their results, HW conjecture that random\-ly distributed
  point masses should cause the largest lensing effects of all
  possible mass distributions for a given average mass density. Note
  that the point-mass case corresponds to the limit in which all
  matter clumps lie well within their Einstein radii [see Eq.~(\ref{eq:rE})]. 
\end{itemize}
{\bf Comment on relevant scales:}\\
The Einstein radius is a characteristic length scale in the lens plane given
by
\begin{equation}
  \label{eq:rE}
  \xi_0=\sqrt{2 R_{\rm Schw}\frac{D_{d}D_{ds}}{D_{s}}},
\end{equation}
where $R_{\rm Schw}=2M$ is the Schwarzschild radius of the lens and 
$D_{ds}$, $D_{d}$ and $D_{s}$ are angular diameter
distances between  lens and source, observer and  lens,
and observer and source, respectively. The corresponding 
length scale in the source plane, $\eta_{0}$, is 
\begin{equation}
  \eta_{0}=\frac{D_{s}}{D_{d}}\xi_{0}=
  \sqrt{2 R_{\rm Schw}\frac{D_{s}D_{ds}}{D_{d}}}.
\end{equation}
  
For lensing to be important, the source size, $l$, has to be 
smaller than the length scale $\eta_{0}$ introduced by a point mass lens, 
$l< \eta_{0}$.
The source size of a type~I supernova is $\sim 10^{15}$ cm at maximum
luminosity. Putting the source at $z=1$ and the lens at $z=0.5$ in a flat
universe with $\Omega_M=0.3$ and the dimensionless Hubble parameter 
$h=0.65$, we obtain 
\begin{equation}
  \label{eq:R}
  M > 10^{-4} M_{\odot},
\end{equation}
i.e., only compact objects with $M > 10^{-4} M_{\odot}$ should 
be of importance through microlensing of supernovae.

\section{The Navarro-Frenk-White distribution}\label{sec:NFW}

The work of HW has as a first treatment 
been concerned with  the $J^\alpha\!_\beta$ expressions for idealised
cases like point
sources; singular, truncated isothermal spheres (SIS); uniform spheres;
and uniform cylinders. However, another often-used matter distribution is the
one based on the results of detailed N-body simulations of 
structure formation given by Navarro et al. (\cite{art:NFW}). 
They parametrize the density profile by 
\begin{equation}
  \label{eq:nfw}
  \rho(r)=\frac{\rho_{\rm crit}\delta_c}
  {(r/R_s)^\gamma[1+(r/R_s)^{1/\delta}]^{(\beta-\gamma)\delta}} ,
\end{equation}
where $\rho_{\rm crit}=3H^2/8\pi$ is the critical density, $\delta_c$
is a dimensionless density parameter and $R_s$ is a characteristic
radius. 

NFW found that halos found in N-body simulations ranging in mass from dwarf 
galaxies to rich galaxy clusters can be fit by Eq.~(\ref{eq:nfw}) with
$\gamma=1$, $\beta=3$, $\delta=1$.  The potential for this density profile 
is given by
\begin{equation}
  \Phi(r)=-4\pi\rho_{\rm crit}\delta_cR_s^2\frac{\ln(1+x)}{x}
  + {\rm const.} ,
\end{equation}
where $x=r/R_{s}$. 
The matrix $J^\alpha\!_\beta$ can then be obtained analytically, see
\ref{app:JNFW}.

The mass inside radius $r$ of a NFW halo is given by
\begin{equation}
        M(r) = 4\pi\rho_{\rm crit}\delta_{c}R_{s}^{3}
        \left[\ln (1+x) - \frac{x}{1+x}\right],
\end{equation}
Since we want to have the average mass density 
in each cell equal to that of the background cosmology, we get the relation
\begin{equation}
        M = \frac{4\pi}{3}\Omega_M\rho_{\rm crit}R_{c}^{3}.
\end{equation}
Setting $M = M(R_{c})$, we obtain
\begin{equation}
        \delta_{c}=\frac{\Omega_M}{3}
        \frac{x_{c}^{3}}{\left[\ln (1+x_{c}) -
        \frac{x_{c}}{1+x_{c}}\right]}, 
\end{equation}
where $x_{c}=R_{c}/R_{s}$.
That is, for a given mass $M$, $\delta_{c}$ is a function of $R_{s}$.
From the numerical simulations of NFW we also get a relation between 
$\delta_{c}$ and $R_{s}$. This relation is computed 
numerically by a slight modification of a  
{\sc Fortran} routine kindly supplied by 
Julio Navarro. 
%
%
Of course, one wants to find an $R_{s}$ which gives both the required
average density in each cell and compatibility with
the numerical results of NFW.

Generally, $R_{s}$ will be a function of mass $M$, the Hubble parameter 
$h$, $\Omega_M$, $\Omega_{\Lambda}$ and $z$. However, we  use the result 
from Del Popolo (\cite{DP99}) and Bullock et al. (\cite{BU99}) that
$R_{s}$ is approximately constant with redshift. 

A comparison of luminosity distributions between point masses,
isothermal sphe\-res (SIS) and the NFW matter distribution is
presented in Sect. \ref{sec:results}.

\section{Mass distribution}\label{sec:massdist}

As discussed above, HW have shown that when studying luminosity distributions 
and other 
quantities where the null geodesics only differ by infinitesimal amounts,
their method is insensitive to the individual masses and clustering of 
point mass lenses.
For point mass lenses, the mass distribution and number density will 
therefore be of small importance. 
However, realistic modeling of galaxies does call for realistic
mass distributions and number densities, i.e., one has to allow for 
the possibility of $R_c$ to reflect the actual distances between galaxies,
and for galaxy masses to vary in agreement with observations.

One of the advantages of the method of HW is that any mass distribution
and number density, including possible redshift dependencies,
can easily be implemented and used as long as the 
average energy density in each cell agrees with the underlying
FL model.

\begin{figure}
  \resizebox{\hsize}{!}{\includegraphics{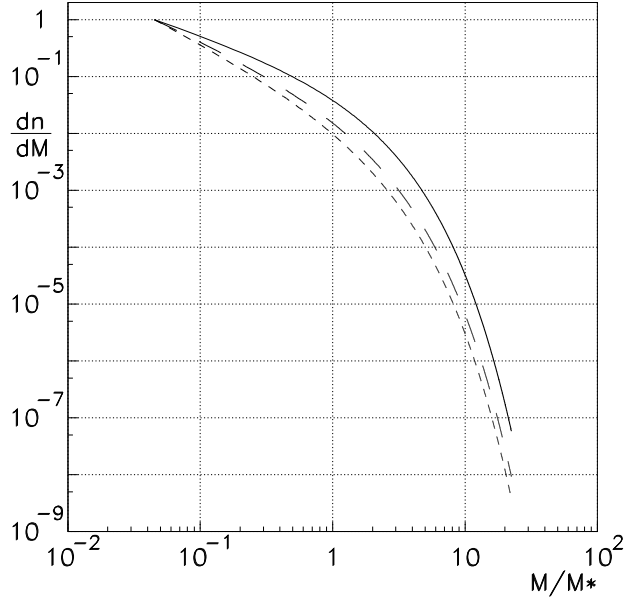}}
  \caption{$dn/dM$ in arbitrary units for $\beta=0.2$ in the interval
    $0.045\leq M/M_*\leq22.4$. The full curve corresponds to
    $\alpha=-0.70$; the long-dashed to $\alpha=-1.07$; and the
    short-dashed to $\alpha=-1.25$.}\label{fig:mdistr} 
\end{figure}

To exemplify this, we derive a galaxy mass distribution, $dn/dM$, 
by combining the Schechter luminosity
function (see, e.g., Peebles \cite{book:Peebles}, Eq.~5.129) 
\begin{eqnarray}
  dn&=&\phi(y)dy , \nonumber \\
  \phi(y)&=&\phi_* y^\alpha e^{-y} , \label{eq:Schechter} \\
  y&=&\frac{L}{L_*} \nonumber ,
\end{eqnarray}
with the mass-to-luminosity ratio
(see, e.g., Peebles \cite{book:Peebles}, Eq.~3.39) 
\begin{equation}
  \label{eq:masstolum}
  \frac{M}{L}\propto M^\beta.
\end{equation}
Normalizing to a ``characteristic'' galaxy with
$L=L_*$ and $M=M_*$, this can be written
\begin{equation}
  \frac{M}{M_*}=y^{1/(1-\beta)} .
\end{equation}
Using Eq.~(\ref{eq:Schechter}), we find that
\begin{eqnarray}
  \frac{dn}{dM}&\propto&y^\delta e^{-y} , \\
  \delta&=&\alpha-\frac{\beta}{1-\beta} .
\end{eqnarray}
In Fig.~\ref{fig:mdistr}, mass distributions for various values of
$\alpha$ are displayed.
Assuming that the entire mass of the universe resides in galaxy 
halos\footnote{In our model with cells joining each other along
the paths of rays, this means that the underdensity in between
galaxies is parametrized by taking the halo mass distribution to
be valid out
to large radii. A discontinuity of mass density then occurs at
the boundary between cells. This has a negligible effect on the
lensing properties, however, since the mass density at the 
discontinuity is very small. A similar, and in fact larger
discontinuity occurs for the truncated isothermal sphere where the
density suddenly drops 
from typically $\rho\sim 10 \rho_{{\rm crit}}$ to $\rho =0$
at the truncation radius.}
we can write 
\begin{equation}
        \int_{y_{\rm min}}^{y_{\rm max}}n(y)M(y)dy = \rho_{m}.
\end{equation}
Using the Schechter luminosity function
and the mass-to-luminosity ratio we get
\begin{equation}
  \label{eq:mref}
        M_{\ast} = \frac{\Omega_M\rho_{\rm crit}}{n_{\ast}
          \int_{y_{\rm min}}^{y_{\rm max}}
          y^{\alpha+\frac{1}{1-\beta}}e^{-y}dy}.
\end{equation}
Thus, by supplying values for $n_{\ast}$, reasonably well-det\-ermined 
by observations, and $y_{\rm min}$ and $ y_{\rm max}$, on which the
dependence of $M_{\ast}$ is weak, together with parameters $\alpha$
and $\beta$ we can obtain an $M_{\ast}$ consistent with $\Omega_M$.

Note that assuming that the Schechter luminosity function and the 
mass-to-luminosity ratio is valid over the entire luminosity range 
(i.e., putting $y_{\rm min}=0$ and $y_{\rm max}=\infty$), we get
\begin{equation}
  \label{eq:mref2}
  M_{\ast} = \frac{\Omega_M\rho_{\rm crit}}{n_{\ast}\Gamma (p+1)},
\end{equation}
where $p=\alpha + 1/(1-\beta),\;p>-1$.

A compilation of parameter values can be found in \ref{app:parametervalues}.
For $0.5<y<2$, $\alpha =-0.7$ and $\beta =0.2$, Eq.~(\ref{eq:mref2}) can be 
written
\begin{equation}
  \label{eq:mrefestimate}
  M_{\ast}\simeq 7.5\cdot 10^{13}\,\Omega_M\left(\frac{0.65}{h}\right)
  \left(\frac{1.9\cdot 10^{-2}h^3}{n_{\ast}\,{\rm Mpc}^3}\right)M_{\odot}.
\end{equation}

\subsection{Truncation radii for SIS-lenses}\label{sec:SIScutoff}

The SIS model is built on the assumption that stars (and other mass 
components) behave like particles in an ideal gas, confined by their 
spherically symmetric gravitational potential. The density profile is 
given by
\begin{equation}
  \rho_{{\rm SIS}}(r)=\frac{v^{2}}{2\pi }\frac{1}{r^{2}},
\end{equation}
where $v$ is the line-of-sight velocity dispersion of the mass particles.
[Note that this corresponds to Eq.~(\ref{eq:nfw}) with $\beta =\gamma =2$,
$\delta\neq 0$ and $v^{2}/(2\pi )=\rho_{{\rm crit}}\delta_{c}R_{s}^{2}$.]
The mass of a SIS halo truncated at radius $d$ is then given by
\begin{equation}
        M(d) = \int_0^r \rho\,(r)dV =2 v^{2}d.
\end{equation}
We want this to be equal to the mass given by the Schechter distribution $M$,
\begin{equation}
        2 v^{2}d=M
        \quad\Rightarrow\quad d=\frac{M_{*}}{2 v_{*}^{2}}
        \left (\frac{M}{M_{*}}\right )\left (\frac{v}{v_{*}}\right )^{-2},
        \label{eq:d}
\end{equation}
where, in addition to $M_{*}$,  we have introduced a characteristic velocity
dispersion $v_{*}$. Combining the Faber-Jackson relation
\begin{equation}
  \frac{v}{v_{*}}=y^{\lambda} ,
\end{equation}
where $y$ is defined in Eq.~(\ref{eq:Schechter}), with the
mass-to-luminosity ratio, Eq.~(\ref{eq:masstolum}), we can substitute
for $v$ in Eq.~(\ref{eq:d}), 
\begin{equation}\label{eq:dcalc}
  d=\frac{M_{*}}{2 v_{*}^{2}}\; 
  \left (\frac{M}{M_{*}}\right )^{1-2\lambda (1-\beta)}
\end{equation}
For values of $v_{*}$ and $\lambda$, see \ref{app:parametervalues}. 
Using Eq.~(\ref{eq:mrefestimate}), we can write the truncation radius for a 
halo with mass $M=M_{\ast}$ as
\begin{equation}
  \label{eq:destimate}
  d\simeq 3.3\,\Omega_M\left(\frac{0.65}{h}\right)
  \left(\frac{1.9\cdot 10^{-2}h^3}{n_{\ast}\,{\rm Mpc}^3}\right)\mbox{Mpc}.
\end{equation}

\begin{figure}
  \resizebox{\hsize}{!}{\includegraphics{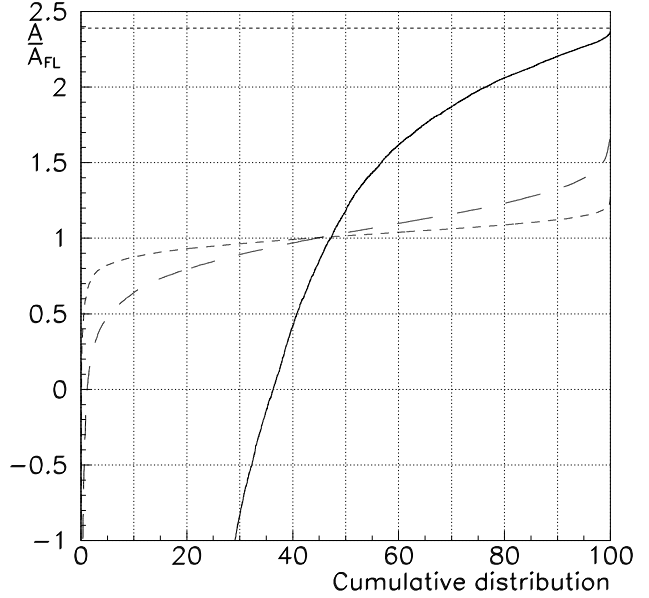}}
  \caption{Beam areas normalized to $A_{\rm FL}$ for 10\,000 
    sources at redshift $z=3$ in a $\Omega_M=1$, $\Omega_\Lambda=0$
    universe. Negative areas correspond to secondary images. The
    dashed line at the top indicates the empty-beam value $A_{\rm empty}$.
    }\label{fig:iso-nfw} 
\end{figure}

A comparison between point masses and isothermal spheres for sources
at $z=3$ in a $\Omega_M=1$, $\Omega_\Lambda=0$ universe is presented
in Fig.~\ref{fig:iso-nfw}, where the
full line corresponds to matter in the form of point masses. This plot
agrees well with Fig.~5 of HW. The long-dashed line corresponds to
matter clumped in isothermal spheres with cut-off radius $d=200$
kpc. We choose this fix $d$ in order to be able to compare directly
with Fig.~11 of HW. Both lines agree very well with the results in HW. 
Finally, the short-dashed line corresponds to isothermal
spheres for which the cut-off radius $d$ has been calculated according
to Eq.~(\ref{eq:dcalc}) (see Sect. \ref{sec:SIScutoff}). For the
presented model, this results in a typical cut-off of $d\sim 3$ Mpc,
which explains why this case comes closer to the filled-beam value.

\section{Results}\label{sec:results}

\begin{figure}
  \resizebox{\hsize}{!}{\includegraphics{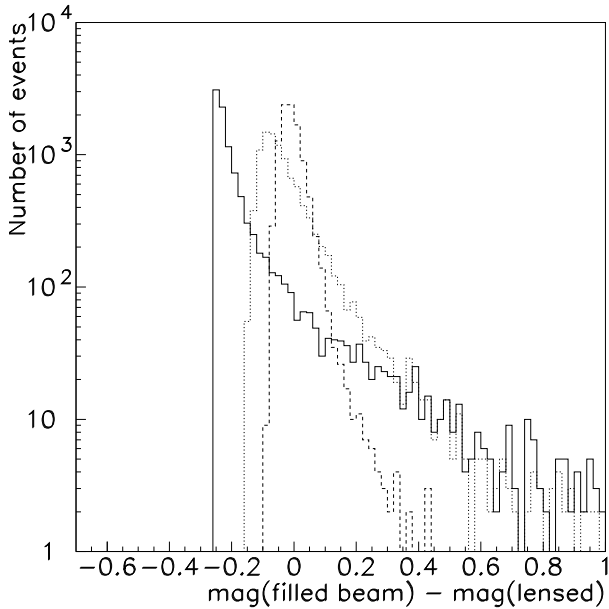}}
  \caption{Luminosity distributions for 10\,000 sources at
    redshift $z=1$ in a $\Omega_M=1$, $\Omega_\Lambda=0$
    universe. The magnification zero point is the filled-beam value.
    The full line corresponds to the point-mass case; the
    dashed line is the distribution for SIS halos, and the dotted line
    is the NFW case. This plot can be compared with Figs.~18 and 20 of 
    HW.}\label{fig:model-1} 
\end{figure}

\begin{figure}
  \resizebox{\hsize}{!}{\includegraphics{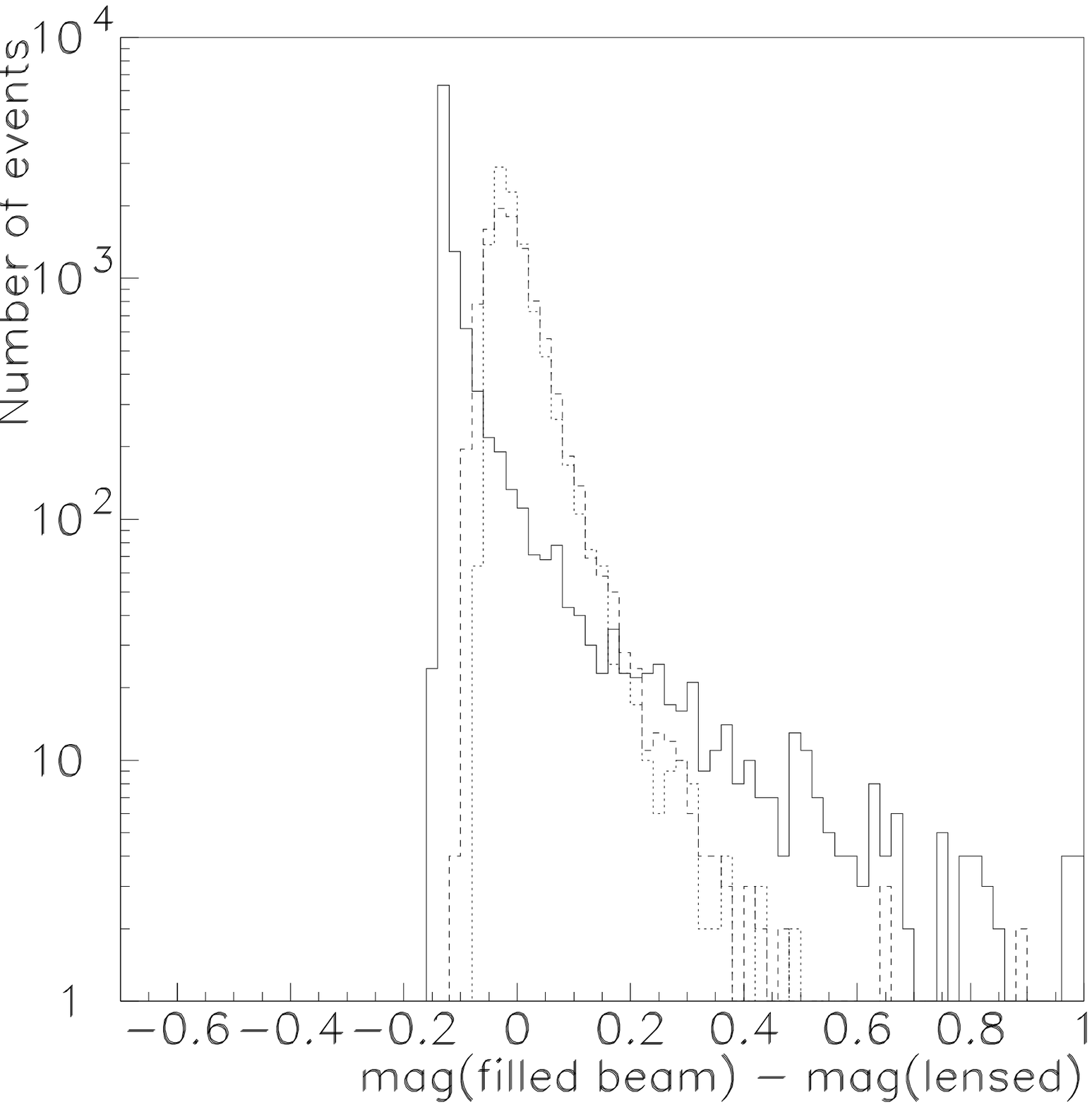}}
  \caption{Luminosity distributions for 10\,000 sources at
    redshift $z=1$ in a $\Omega_M=0.3$, $\Omega_\Lambda=0.7$
    universe. The magnification zero point is the filled-beam value.
    The full line corresponds to the point-mass case; the
    dashed line is the distribution for SIS halos, and the dotted line
    is the NFW case. This plot can be compared with Fig.~22 of 
    HW.}\label{fig:model-2} 
\end{figure}

\begin{figure}
  \resizebox{\hsize}{!}{\includegraphics{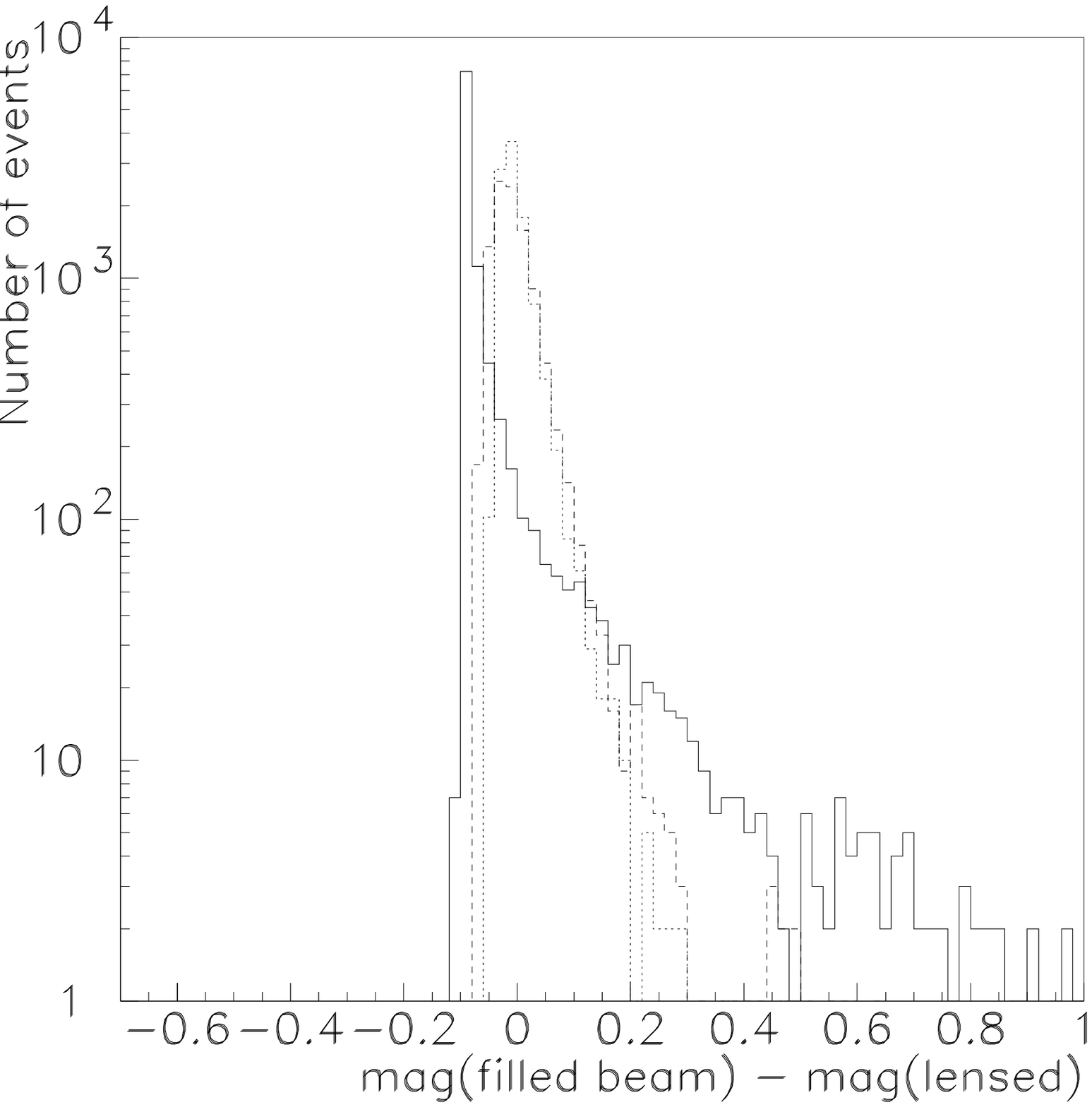}}
  \caption{Luminosity distributions for 10\,000 sources at
    redshift $z=1$ in a $\Omega_M=0.3$, $\Omega_\Lambda=0$
    universe. The magnification zero point is the filled-beam value.
    The full line corresponds to the point-mass case; the
    dashed line is the distribution for SIS halos, and the dotted line
    is the NFW case.}\label{fig:model-3} 
\end{figure}

\begin{figure}
\resizebox{\hsize}{!}{\includegraphics{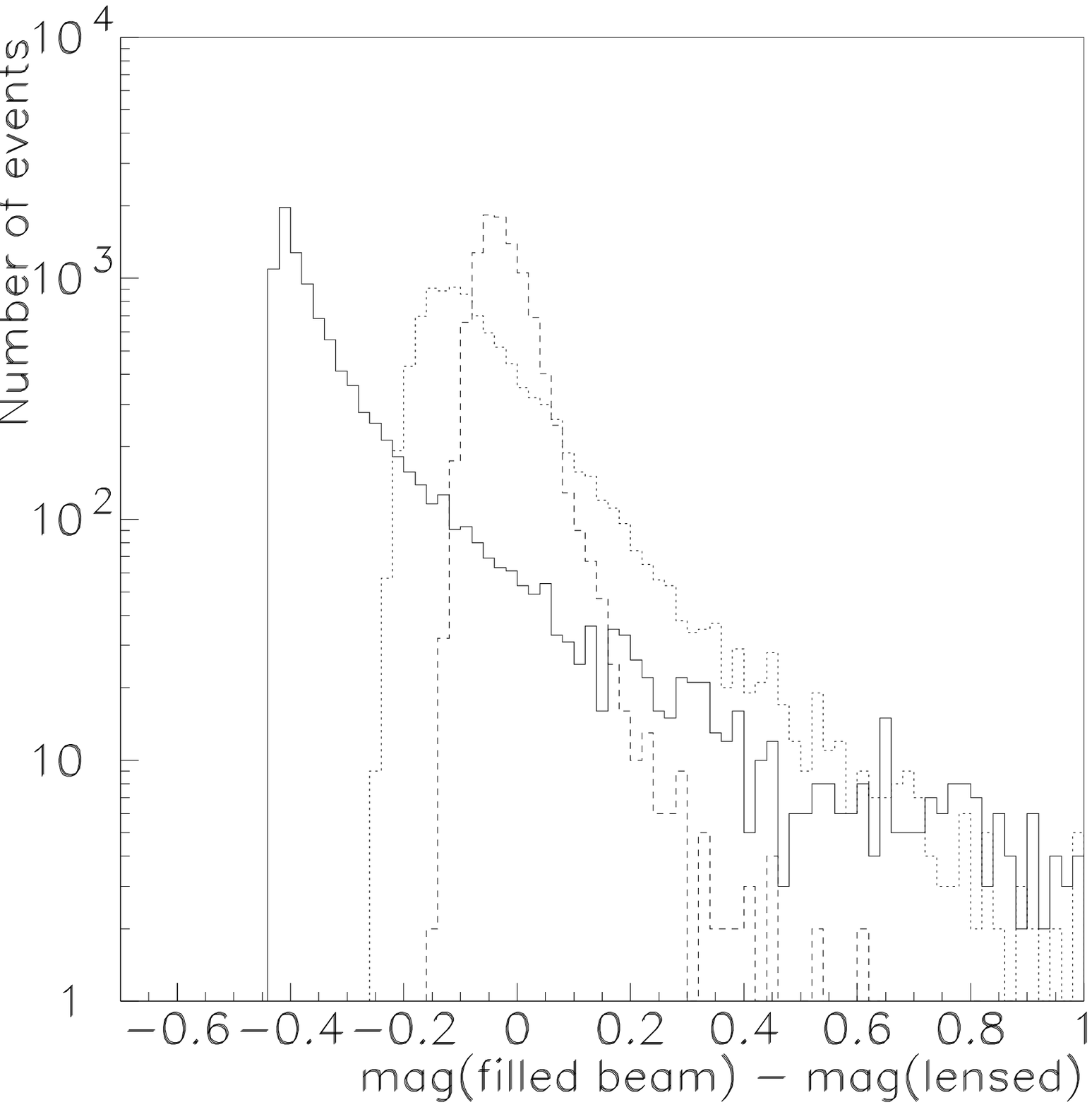}}
  \caption{Luminosity distributions for 10\,000 sources at
    redshift $z=1.5$ in a $\Omega_M=1$, $\Omega_\Lambda=0$
    universe. The magnification zero point is the filled-beam value.
    The full line corresponds to the point-mass case; the
    dashed line is the distribution for SIS halos, and the dotted line
    is the NFW case.}\label{fig:model-4} 
\end{figure}

\begin{figure}
\resizebox{\hsize}{!}{\includegraphics{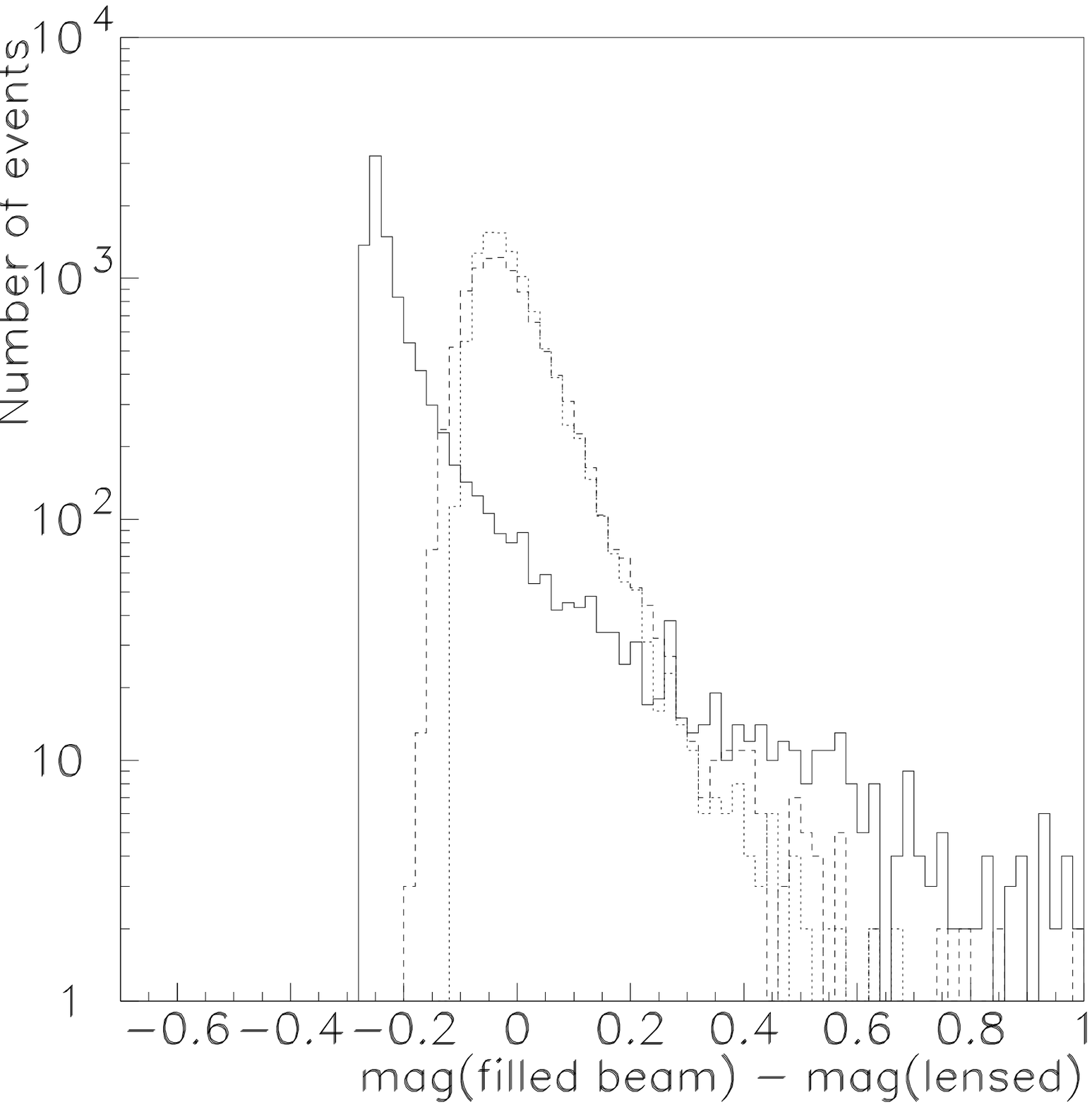}}
  \caption{Luminosity distributions for 10\,000 sources at
    redshift $z=1.5$ in a $\Omega_M=0.3$, $\Omega_\Lambda=0.7$
    universe. The magnification zero point is the filled-beam value.
    The full line corresponds to the point-mass case; the
    dashed line is the distribution for SIS halos, and the dotted line
    is the NFW case.}\label{fig:model-5} 
\end{figure}

\begin{figure}
\resizebox{\hsize}{!}{\includegraphics{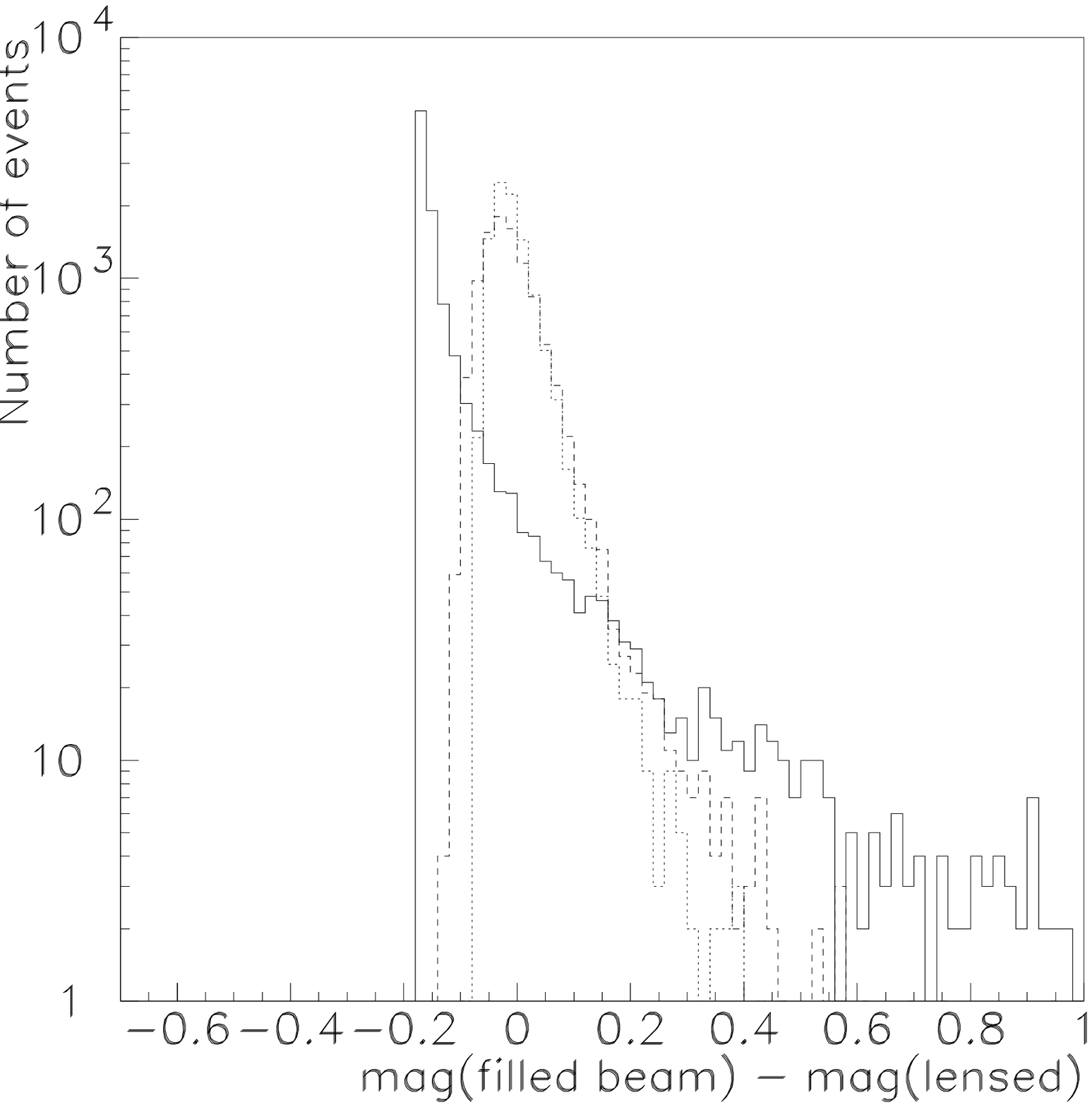}}
  \caption{Luminosity distributions for 10\,000 sources at
    redshift $z=1.5$ in a $\Omega_M=0.3$, $\Omega_\Lambda=0$
    universe. The magnification zero point is the filled-beam value.
    The full line corresponds to the point-mass case; the
    dashed line is the distribution for SIS halos, and the dotted line
    is the NFW case.}\label{fig:model-6} 
\end{figure}

\begin{figure}
\resizebox{\hsize}{!}{\includegraphics{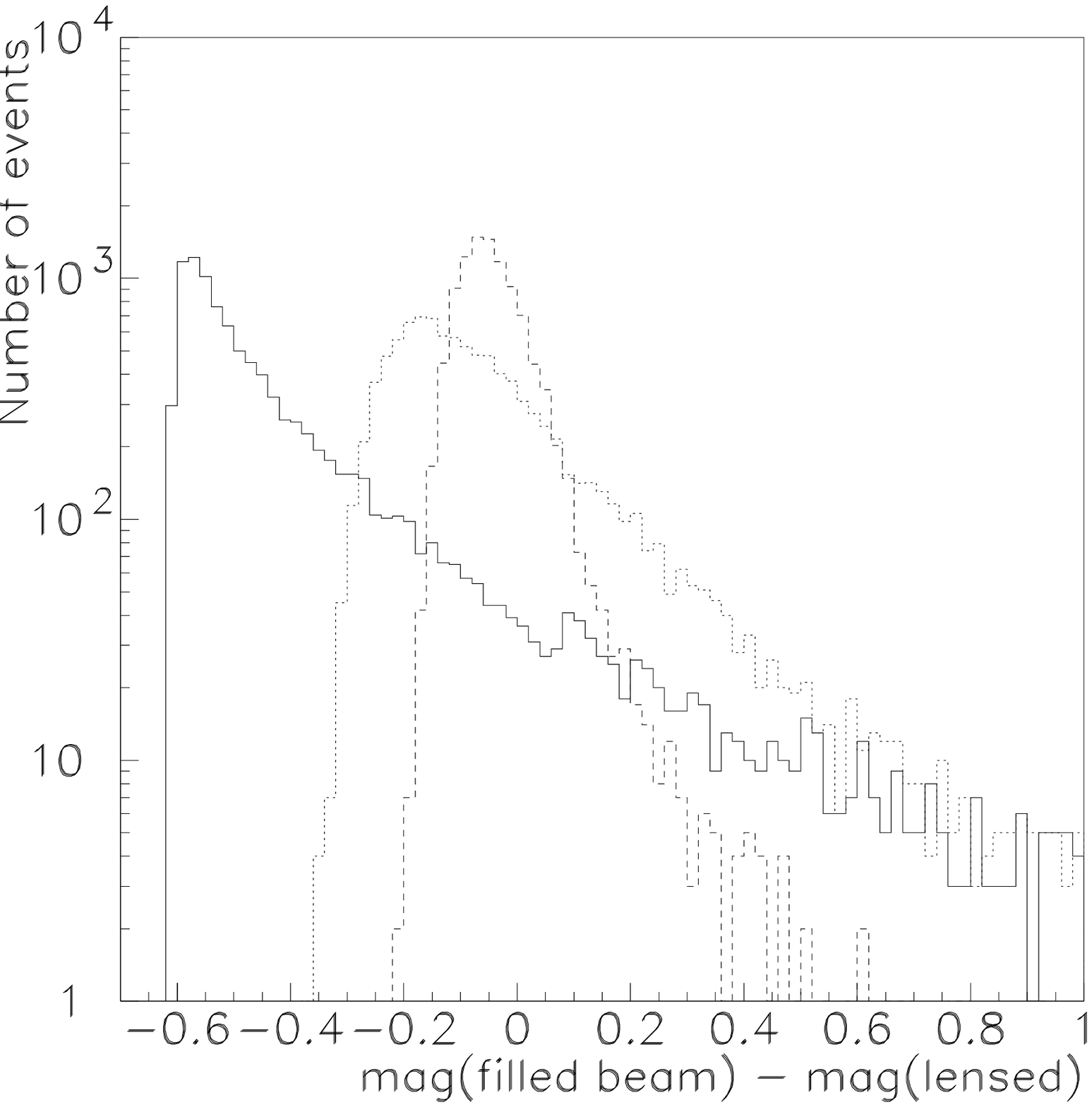}}
  \caption{Luminosity distributions for 10\,000 sources at
    redshift $z=2$ in a $\Omega_M=1$, $\Omega_\Lambda=0$
    universe. The magnification zero point is the filled-beam value.
    The full line corresponds to the point-mass case; the
    dashed line is the distribution for SIS halos, and the dotted line
    is the NFW case.}\label{fig:model-7} 
\end{figure}

\begin{figure}
\resizebox{\hsize}{!}{\includegraphics{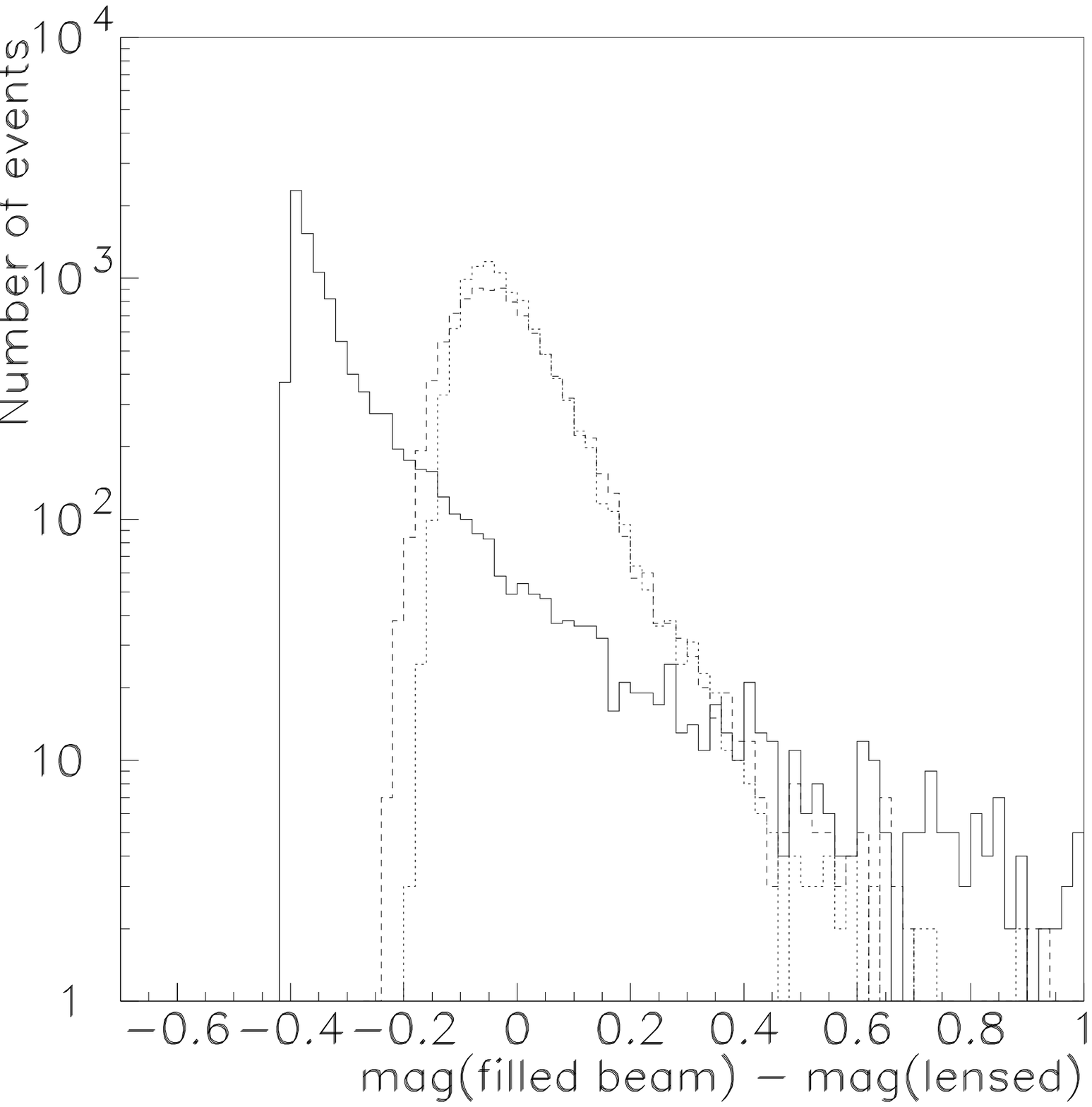}}
  \caption{Luminosity distributions for 10\,000 sources at
    redshift $z=2$ in a $\Omega_M=0.3$, $\Omega_\Lambda=0.7$
    universe. The magnification zero point is the filled-beam value.
    The full line corresponds to the point-mass case; the
    dashed line is the distribution for SIS halos, and the dotted line
    is the NFW case.}\label{fig:model-8} 
\end{figure}

\begin{figure}
\resizebox{\hsize}{!}{\includegraphics{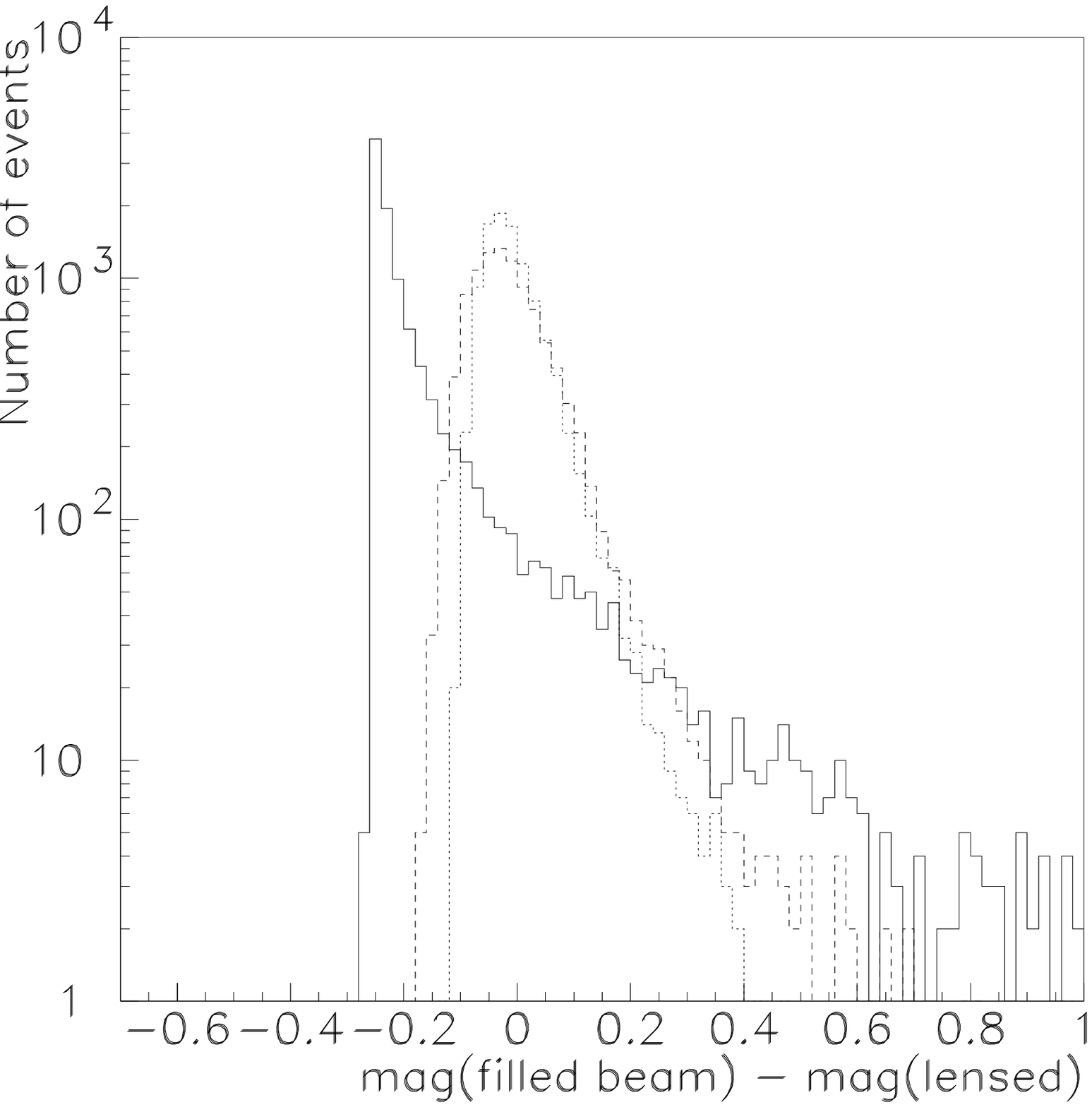}}
  \caption{Luminosity distributions for 10\,000 sources at
    redshift $z=2$ in a $\Omega_M=0.3$, $\Omega_\Lambda=0$
    universe. The magnification zero point is the filled-beam value.
    The full line corresponds to the point-mass case; the
    dashed line is the distribution for SIS halos, and the dotted line
    is the NFW case.}\label{fig:model-9} 
\end{figure}

In Figs.~\ref{fig:model-1}--\ref{fig:model-9},
we compare the luminosity distributions obtained for NFW halos with the
point-mass and SIS matter distributions.
Comparing Fig.~\ref{fig:model-1} with Fig.~18 and 20 of HW, we see that 
the point-mass cases agree very well. The SIS luminosity 
distributions differs in that our results is shifted towards the filled
beam value. This is due to the fact that the truncation radii used 
[see Eq.~(\ref{eq:destimate})] are an order of magnitude larger than the 
(constant) value of $d=200$ kpc used in HW.

Lensing effects are governed by the lensing potential of the galaxy model 
used, and the number of lenses between the source and the observer.
As noted in the discussion of Fig.~\ref{fig:Xmatter}, 
a larger number of lenses will
tend to make lensing effects for a given galaxy model more prominent. 
For a fixed source redshift and a fixed number density of lenses,
the number of lenses will only be a function of the cosmology. 
We get the largest 
number of lenses in the $\Omega_M =0.3,\,\Omega_{\Lambda}=0.7$ cosmology,
followed by the $\Omega_M =0.3,\,\Omega_{\Lambda}=0$ case, and the smallest
number of lenses for $\Omega_M =1,\,\Omega_{\Lambda}=0$.

For point masses, the lensing potential will be unaffected by the cosmology.
As discussed in Sect.~\ref{sec:HWsummary}, point masses cause the largest 
lensing effects for a given average mass density.

\begin{figure}
\resizebox{\hsize}{!}{\includegraphics{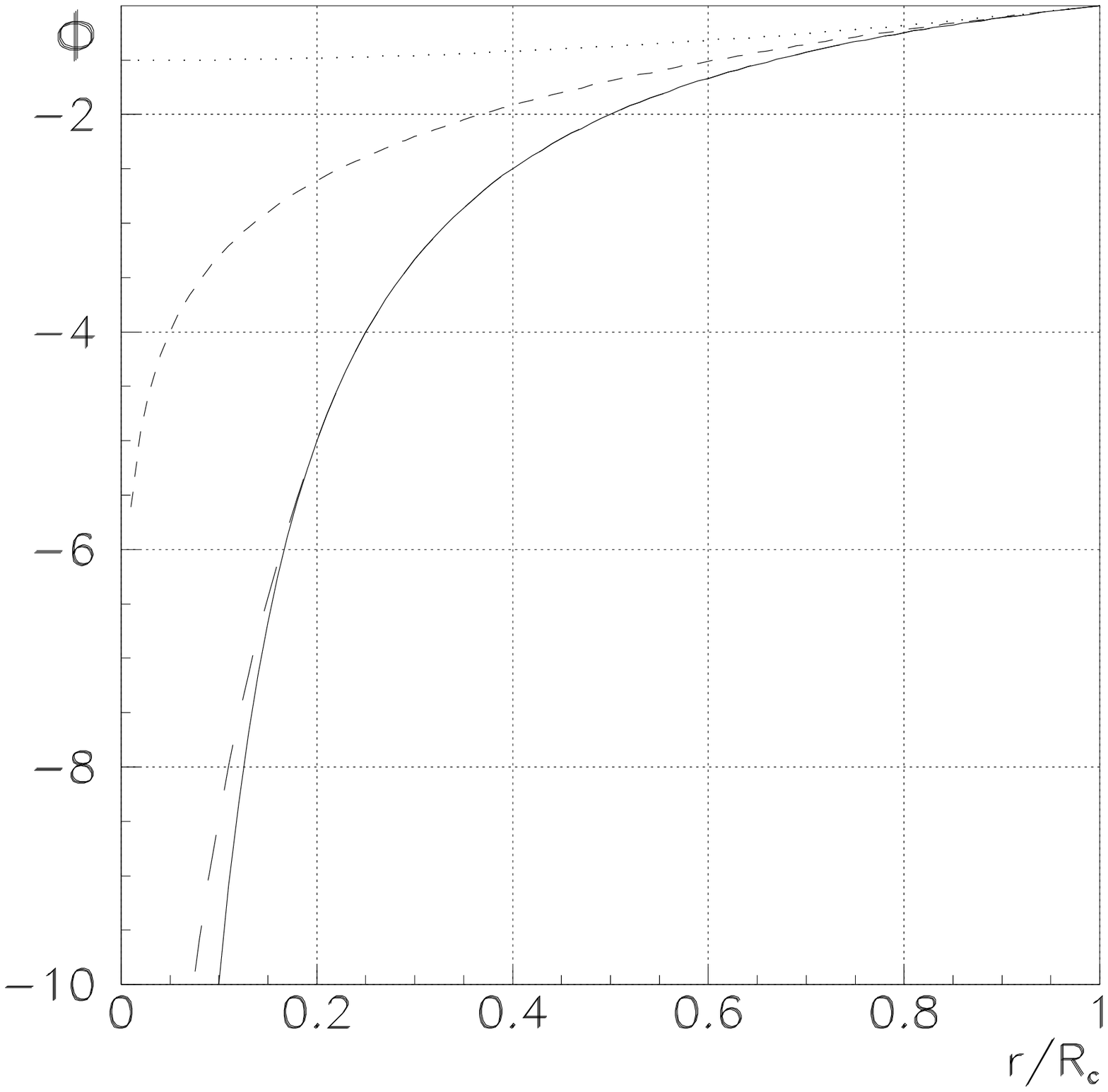}}
  \caption{Lensing potentials normalized to $M=1$. The dotted line
corresponds to the potential of homogeneously distributed matter
(filled-beam);  the short--dashed line is a SIS halo with $d=R_c$;
the long--dashed line is a SIS halo with $d=0.2R_c$, and the full line
corresponds to the point-mass case.}\label{fig:sispotentials}
\end{figure}

For a SIS halo, the truncation radius, $d$, is proportional to $\Omega_M$ 
[see Eq.~(\ref{eq:mref}) and Eq.~(\ref{eq:dcalc})]. 
Since the potential for a truncated SIS halo is equal to the point-mass 
potential for $r\ge d$ (see Fig.~\ref{fig:sispotentials}), 
the SIS luminosity distribution will be 
shifted towards the point-mass distribution when lowering $\Omega_M$.
In fact, the agreement between our results and those in HW is better for 
lower $\Omega_M$, as can be seen
by comparing the curve corresponding to the SIS luminosity distribution in 
Fig.~\ref{fig:model-2} with Fig.~22 of HW. 
Taking also the number of lenses into consideration, we expect the largest
lensing effects for the $\Omega_M =0.3,\,\Omega_{\Lambda}=0.7$ cosmology,
followed by the $\Omega_M =0.3,\,\Omega_{\Lambda}=0$ case, and the smallest
effects in the $\Omega_M =1,\,\Omega_{\Lambda}=0$ cosmology. 

\begin{figure}
\resizebox{\hsize}{!}{\includegraphics{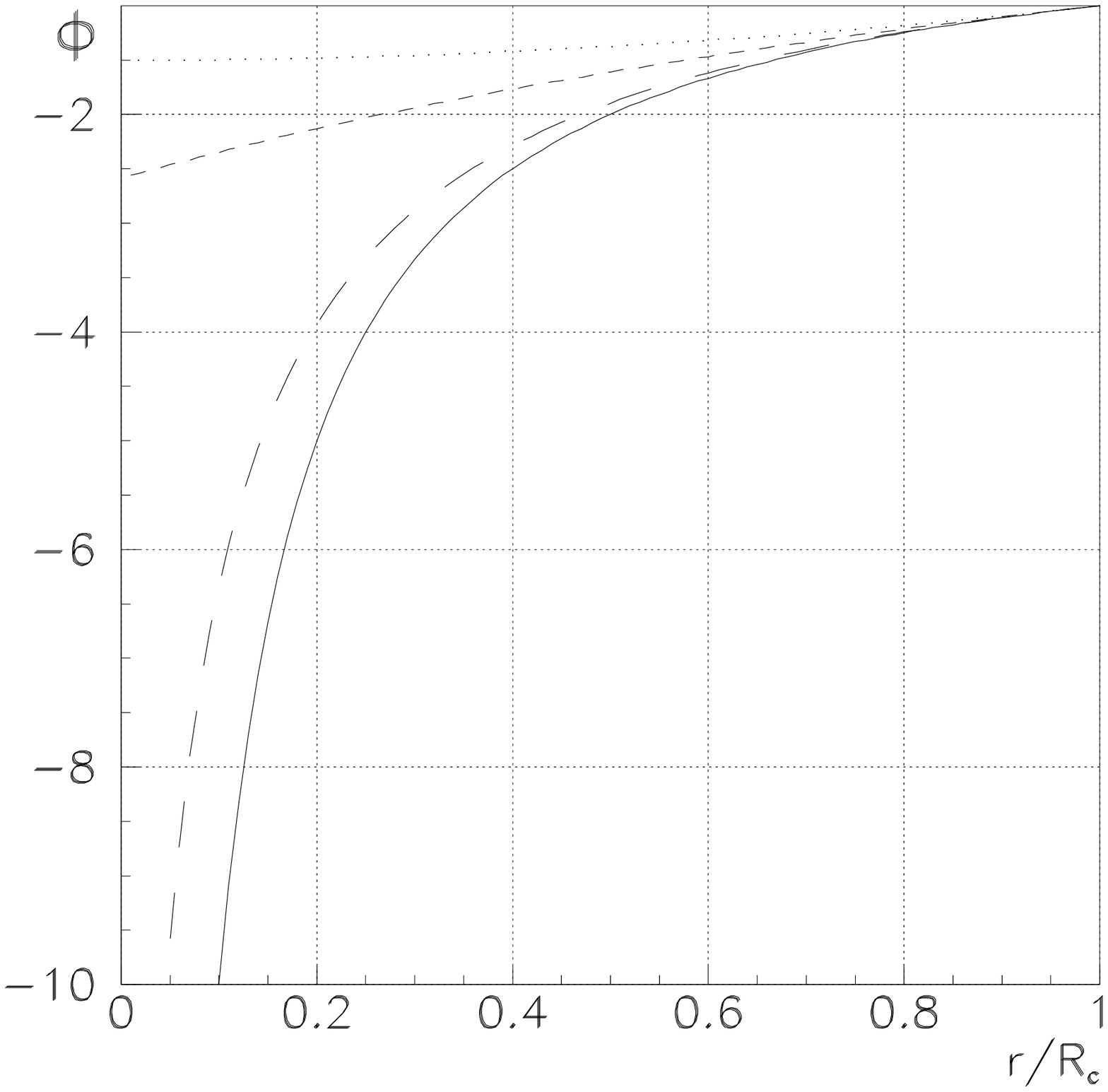}}
  \caption{Lensing potentials normalized to $M=1$. The dotted line
corresponds to the potential of homogeneously distributed matter
(filled-beam);  the short--dashed line is a NFW halo with $R_s=R_c$;
the long--dashed line is a NFW halo with $R_s=0.01R_c$, and the full line
corresponds to the point-mass case.}\label{fig:nfwpotentials}
\end{figure}

For NFW halos, the value of the characteristic radius $R_s$ will depend
on the cosmology used, the largest value obtained with
the $\Omega_M =0.3,\,\Omega_{\Lambda}=0$ cosmology,
followed by $\Omega_M =0.3,\,\Omega_{\Lambda}=0.7$ and
the $\Omega_M =1,\,\Omega_{\Lambda}=0$ case.
Now, it can be shown that for a fixed halo mass, $M$, the NFW potential will 
approach the point-mass potential when lowering $R_s$, 
see Fig.~\ref{fig:nfwpotentials}\,\footnote{In fact, in the limit 
$R_s\to 0$, $\Phi_{\rm NFW}(r)\to \Phi_{\rm point}(r)$.}.
Therefore, we expect the closest resemblance to the point-mass case for 
$\Omega_M =1,\,\Omega_{\Lambda}=0$, followed by 
$\Omega_M =0.3,\,\Omega_{\Lambda}=0.7$ and 
$\Omega_M =0.3,\,\Omega_{\Lambda}=0$.

\section{Limitations of method}\label{sec:limit}

\begin{figure}
  \resizebox{\hsize}{!}{\includegraphics{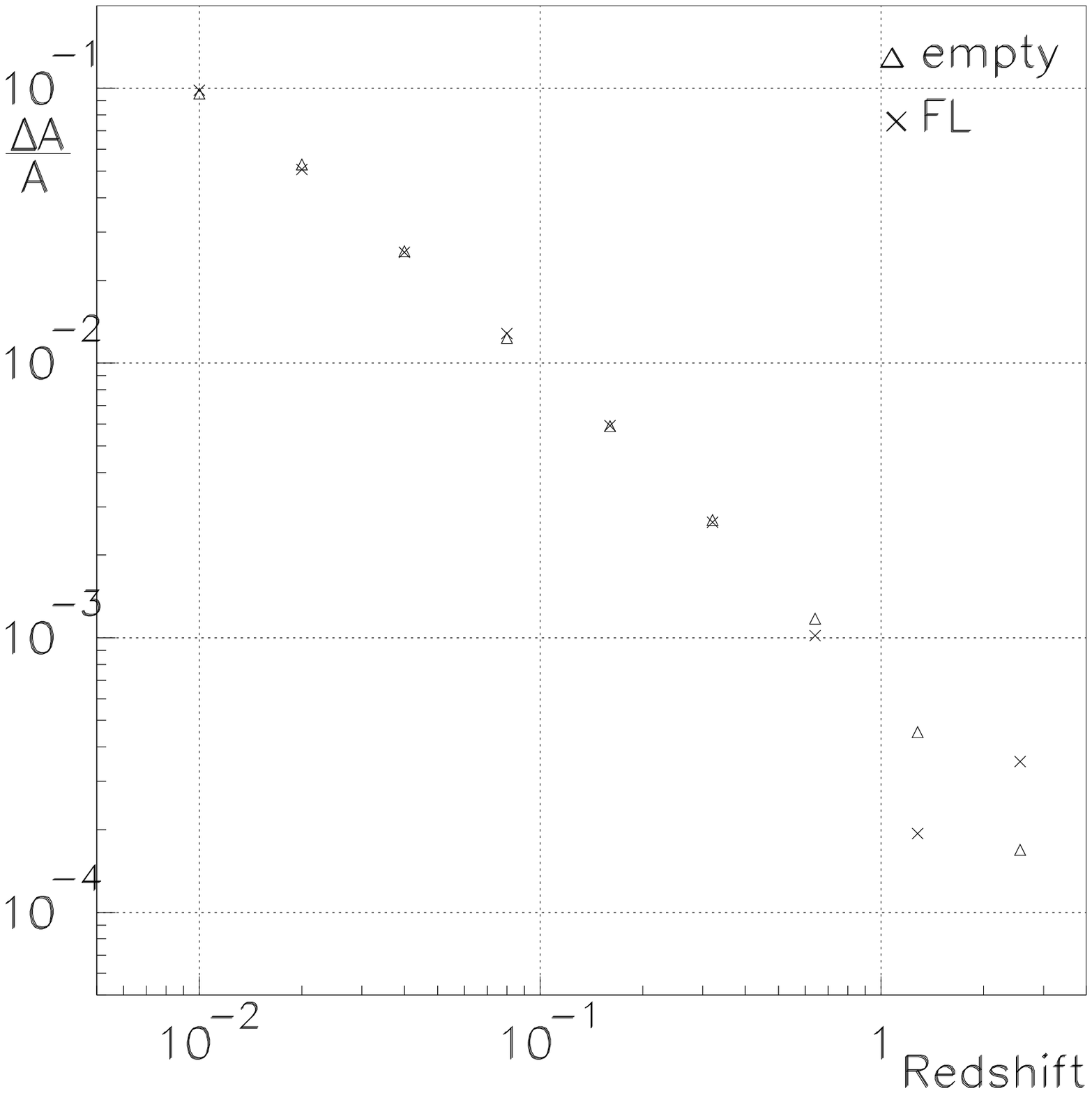}}
  \caption{Fractional mean error in area vs redshift.}\label{fig:error} 
\end{figure}
The method of HW, or rather the numerical implementation of it, has some 
limitations due to the approximations used. 
Since the area of the light beam when exiting a cell is computed by linear 
interpolation from the value before entering the cell 
[see Eq.~(\ref{eq:AdA})], we need very small cells or very many cells to
get robust results. Small cells means shorter interpolations while a
large number of cells tends to average out the error in every 
interpolation. 

In order to quantify this effect we have made some tests 
with  empty and filled beams using cell sizes obtained from the 
Schechter matter distribution. In Fig.~\ref{fig:error}, we have plotted 
the mean fractional error in area vs. redshift.
To get a mean fractional error less than one percent, one would need to
have $z>0.1$. Note that this is the average error of individual
measurements; the average value will be very close to the real value even at
very small redshifts. For the cosmology used ($\Omega_M=0.3, 
\Omega_{\Lambda}=0.7$), a redshift of $\sim 0.1$
corresponds to $\sim 60$ cells.

\section{Multiple imaging}\label{sec:mi}

Another drawback with the HW method is that there is no way to keep track
of multiple images from the same source. This poses no problem when the
time delay between a primary image and secondary images is big enough
to allow observational separation of the images. 
However, if image separation is not observationally feasible, we have
to compensate for multiple images. This will be the case when
considering microlensing, see Sect.~(\ref{subsec.ola}).

\begin{figure}[htp]
  \begin{center}
    \resizebox{\hsize}{!}{\includegraphics{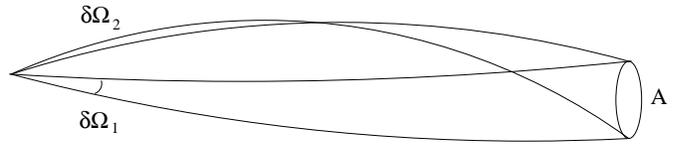}}
    \caption{Multiple imaging.}
    \label{fig:focusing4}
  \end{center}
\end{figure}

A typical case of multiple imaging is shown in Fig.~\ref{fig:focusing4}.
Note that the secondary image has 
undergone a caustic crossing and that its parity has been reversed (this 
corresponds to $A<0$). 
Also, the secondary image is 
demagnified compared to the primary image (remember that it is the 
relation between $\delta\Omega$ and the area that determines the 
magnification). Since the quantity $\delta\Omega$ is more or less constant 
in our method, areas of secondary images often becomes very big, i.e., 
very dim.

For a point mass lens, the time delay between two images with flux ratio
$r$ is given by
\begin{equation}
  \Delta t=2R_{\rm Schw}(1+z_L)\tau (r),
\end{equation}
where
\begin{equation}
  \tau (r)=\frac{1}{2}(\sqrt{r}-\frac{1}{\sqrt{r}}+\ln{r}).
\end{equation}
For a lens at redshift $z=0.5$ and a flux ratio of two, we get
\begin{equation}
  \Delta t \sim 2\cdot 10^{-5}\left(\frac{M}{M_{\odot}}\right)\, s.
\end{equation}
As an illustration, to get a time delay of the order of weeks, we need 
$M\sim 10^{11}M_{\odot}$.

The image separation in the same case (assuming a source at $z=1$)
is given by
\begin{equation}
  \Delta\theta \sim 3\cdot 10^{-6}\left(\frac{M}{M_{\odot}}\right) ",
\end{equation}
i.e., a mass of $M\sim 10^{11}M_{\odot}$ will give an image separation of
$\sim 1"$. In the realistic case that a galaxy is causing the lensing,
the surface brightness of the lens may make the detection of secondary images
impossible (Porciani \& Madau \cite{madau}).

\subsection{A point-mass universe}\label{subsec.pmu}

\begin{figure}[htp]
  \begin{center}
    \resizebox{\hsize}{!}{\includegraphics{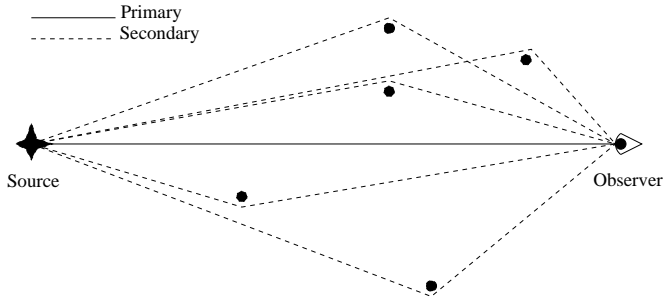}}
    \caption{Point mass universe, first case.}
    \label{fig:points}
  \end{center}
\end{figure}

Now, let us discuss what kind of information we can obtain from 
secondary images. In Fig.~\ref{fig:points} we see what a source would 
look like in a universe filled with point-mass lenses. Since the 
deflection angle from a point mass can be very large, there 
will, apart from the primary image, be one secondary image next to 
every point mass lying in the vicinity of the line-of-sight to the source 
(not to mention all images with two or more caustic crossings).\footnote{The 
deflection angle is given by $\hat\alpha = 2R_{\rm Schw}/\xi$.} 
Most of these images will be \emph{very} dim, 
i.e. not observable. 

\begin{figure}[htp]
  \begin{center}
    \resizebox{\hsize}{!}{\includegraphics{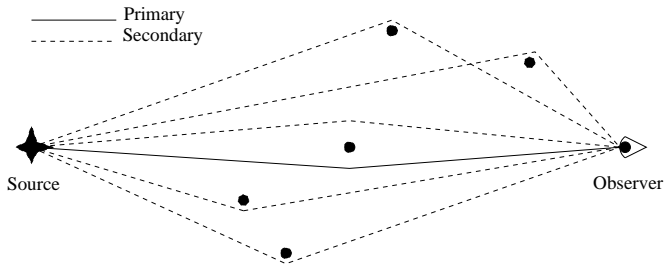}}
    \caption{Point mass universe, second case.}
    \label{fig:points2}
  \end{center}
\end{figure}

In our implementation of the HW method, we can happen to follow any one
 of these. 
However, the vast majority of the flux will almost always be in 
the primary image, the only exception being the case where we have a 
point mass very close to the line-of-sight (see Fig.~\ref{fig:points2}).
In this case, the primary and the brightest secondary image 
will have almost equal flux.
If we want to consider the total flux from a source we need to correct for 
these images. Of course, the correction would have to be larger in the case
of Fig.~\ref{fig:points2}. Multiple-image corrections are discussed in
Sects.~\ref{subsec.ola} and \ref{subsec.olsis} below. 

Note that if we happen to 
follow any of the weak secondary images in Fig.~\ref{fig:points2}, we do not 
have any information on the flux of the primary image, i.e., we would not be
able to discriminate between the case of Fig.~\ref{fig:points} and
Fig.~\ref{fig:points2}. From this we conclude
that weak secondary images does not give any valuable information 
concerning the total flux of a source. 

\subsection{A SIS-galaxy universe}\label{subsec.sisu}

Filling the universe with SIS-type galaxies, the situation would be 
quite different. No longer can we have very large deflection 
angles, and multiple images will therefore be less common.\footnote{For 
SIS lenses, the deflection angle is given by 
$\hat\alpha = 4\pi (v/c)^{2}|\xi|/\xi$, where
$v$ is the velocity dispersion of the lens and $\xi$ is the impact 
parameter, i.e., the magnitude of the deflection does not depend on $\xi$.} 
Since we are typically dealing with galaxy masses, images will be separable 
and we need only to consider secondary images bright enough to be 
observable. Since for (almost) every source, there will be at most one 
secondary image, it is a more straightforward procedure to 
associate to every secondary image a primary image. 
\begin{figure}[htp]
  \begin{center}
    \resizebox{\hsize}{!}{\includegraphics{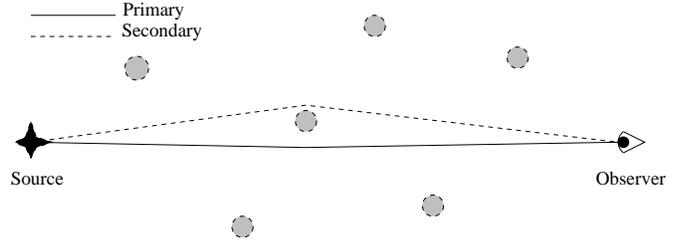}}
    \caption{SIS universe.}
    \label{fig:sis}
  \end{center}
\end{figure}

\subsection{One-lens approximation for point masses}\label{subsec.ola}

Here, we will discuss reasonable corrections when images are 
non-separable. 
Since we have a very large number of images with two or 
more caustic crossings contributing to the flux of the total image, we 
can not correct for each one of them. Instead, we will make a first 
correction for the brightest secondary image using the approximation 
that lensing effects from a single lens (cell) is dominant. This is the
so called one-lens approximation. Possible tests of the validity
of this assumption is discussed in Sect.~\ref{sec:cc}.
Its plausibility is also reinforced by Figs.~8 and 9 in HW.

We can now use standard analytical expressions for one-lens systems 
for this correction: 
\begin{equation}
    \mu_{\rm corr}=\vert 2\mu -1\vert.
    \label{eq:mucorr}
\end{equation}
In fact, this formula is valid also for secondary images but
one then has to include the sign in $\mu$, i.e., $\mu_{2}<0$.
However, Eq.~(\ref{eq:mucorr}) is only valid for the brightest 
secondary image, and one should therefore not apply it to very weak, 
secondary images.
Also note, that to make the correction self-consistent, the magnification 
should be defined as the ratio of the expected area if the most prominent 
cell is empty and the real area.
It is easy to see that the most important effect of this correction is to 
make the high magnification tail more pronounced.

\subsection{One-lens approximation for SIS lenses}\label{subsec.olsis}

It is as straightforward to compute the total magnification using the 
one-lens approximation with a SIS-lens. For a primary image 
we get
\begin{equation}
    \mu_{\rm corr}=2(\mu_{1}-1).
    \label{eq:sismu1}
\end{equation}
For a secondary image we get
\begin{equation}
    \mu_{\rm corr}=2(1-\mu_{2}).
    \label{eq:sismu2}
\end{equation}
Since images are separable in the majority of cases, this is perhaps most 
interesting to use as a consistency check for the total luminosity.
  
\section{Consistency checks}\label{sec:cc}

In order to validate our implementation of the method, we have
performed a number of tests.
\begin{itemize}
  \item {\bf Filled and empty beam distances}\\
  First -- in order to get an independent check -- we have compared 
  results from our implementation of the HW method
  using empty cells and cells with 
  a homogeneous dust component with analytical empty-beam and filled-beam 
  results (using the {\tt Angsiz} routine, see Kayser et al.
  \cite{Kayser1995}). 
  This has been done for a variety of cosmologies and the agreement is 
  good even at very high $z$ (less than 1 \% discrepancy up to $z=10$).  
  \item {\bf Comparison with HW}\\
  Second, we have reproduced most of the plots presented by HW. We have also
  tested that primary images satisfy $\frac{1}{N_{\rm tot}}\sum A_1\geq1$,
  see Fig.~\ref{fig:areasum}.
\begin{figure}
  \resizebox{\hsize}{!}{\includegraphics{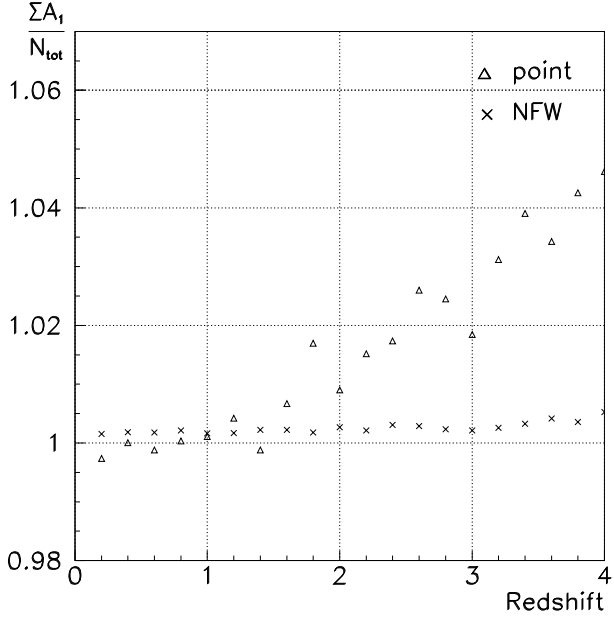}}
  \caption{Sum of primary areas divided by $N_{\rm tot}$ for point masses
    (triangles) and NFW halos (crosses) in a $\Omega_M=0.3$,
    $\Omega_\Lambda=0.7$ universe. The number of primary images for
    each data point is $N_1=20\,000$.}\label{fig:areasum} 
\end{figure}
  \item {\bf Average luminosities}\\
  Imagine a source situated at the center of the sphere in the lower 
  panel of Fig.~\ref{fig:focusing3} and observers sitting on the sphere 
  at some constant redshift. Since the number of photons emitted by the 
  source, as well as the area of the observers sphere is constant, i.e. 
  independent of any details of the distribution of matter as long as the 
  average matter density is the same, observers will on average receive 
  the same flux, regardless of the matter distribution. This average 
  will be equal to the FL value, 
  \begin{equation}
    \langle\mu\rangle = \mu_{\rm FL}.
    \label{eq:mumean}
  \end{equation}
  When computing $\mu$ for a source, one has to consider the total 
  flux if there are multiple images. If we consider primary images only, 
  without any corrections, we would expect to obtain a $\mu$ slightly 
  lower than the FL value (since we lose some flux in secondary images).

  Note that this does not mean that the average area should equal the
  FL area. Instead,
  \begin{eqnarray}
    \langle\mu\rangle =\frac{1}{N}\sum_{i=1}^{N}\frac{A_{\rm empty}}{A_{i}}
    &=&\frac{A_{\rm empty}}{A_{\rm FL}}\Rightarrow \nonumber \\
    \frac{1}{A_{\rm FL}}&=&\frac{1}{N}\sum_{i=1}^{N}\frac{1}{A_{i}} ,
    \label{eq:harmmean}
  \end{eqnarray}
  i.e., the FL area is the harmonic mean of the event areas, not the 
  arithmetic (remember that $\langle A\rangle_{\rm Harm}<\langle 
  A\rangle_{\rm Arit}$). Since we lose some flux in secondary images, we 
  have (considering primary images only)
  \begin{equation}
    A_{\rm FL}<\langle A\rangle_{\rm Harm}<\langle A\rangle_{\rm Arit}.
    \label{eq:a}
  \end{equation}
  See Fig.~\ref{fig:lumtest} for a plot of the average magnification using
  point mass lenses and NFW lenses. The result for SIS halos is not 
  plotted since the result is 
  indistinguishable from the NFW result. 
  Note that since the deviation from 
  the filled-beam value is due to the absence of the flux contained in 
  secondary images, Fig.~\ref{fig:lumtest} indicates that the flux in
  secondary images is very low when using NFW or SIS halos, even at very
  high redshifts.
\begin{figure}
  \resizebox{\hsize}{!}{\includegraphics{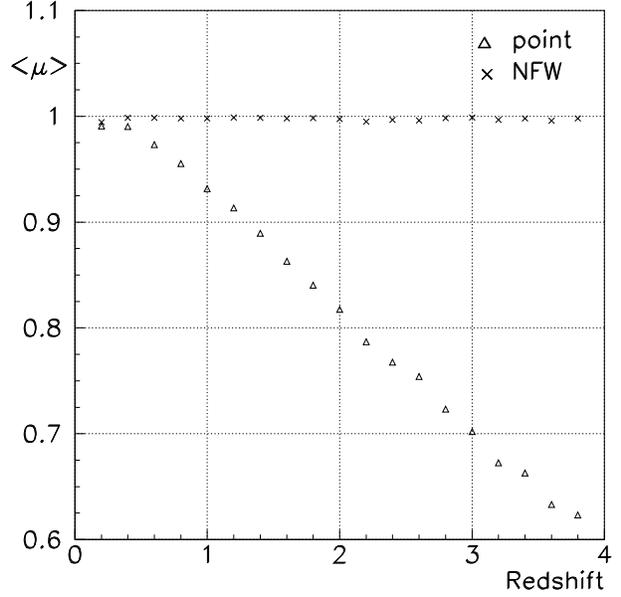}}
  \caption{Average magnification of primary images 
    (normalized to filled-beam values) for point masses 
    (triangles) and NFW halos (crosses) in an $\Omega_M=0.3$,
    $\Omega_\Lambda=0.7$ universe. The number of primary images for
    each data point is $N=20\,000$.}\label{fig:lumtest} 
\end{figure}
  
The relationship between this test and the area test of HW is 
discussed in \ref{app:meanarea}.
  
\end{itemize}

\section{Analytical fitting formulas}\label{sec:formula}

Since the numerical work involved in computing the results presented in
this paper is quite extensive, it may be convenient to use simple
analytical fitting formulas for the probability distributions for the
different cosmologies, which we now give.

The probability distribution of the deviations from filled-beam
magnitudes induced by SIS and NFW lenses,
$\Delta=$mag(filled beam)$-$mag(lensed), can be parametrized by an
exponential function folded with a gaussian distribution: 
\begin{equation}
P(\Delta) \propto \int_{0}^\infty{ e^{-\left(\Delta -\mu -m'\right)^2 
\over 2 \sigma^2 } \cdot e^{-{\left(m' \over \tau\right)}} dm'}
\label{eq:gausexp1}
\end{equation}

Thus, for any combination of cosmological parameters, the corrections
due to gravitational lensing can be described by three parameters: the
mean of the gaussian distribution, $\mu$, the standard deviation
$\sigma$, and the magnification tail, $\tau$. The expression in
Eq.~(\ref{eq:gausexp1}) can be rewritten so that (after normalization)
a simpler integral is invoked:  

\begin{equation}
P(\Delta) = {1 \over 2 \sigma} 
e^{\left({\tau/\sigma - 2(\Delta -\mu) \over 2 \sigma}\right)} 
 \cdot {\rm erfc}\left({\tau/\sigma -(\Delta -\mu) 
\over \sqrt{2 \tau}}\right)
\label{eq:gausexp2}
\end{equation}
where
\begin{equation}
{\rm erfc}(x) = {2 \over \sqrt{\pi}}  \int_x^\infty{e^{-x'^2}  dx'}
\end{equation}

\begin{figure}
  \resizebox{\hsize}{!}{\includegraphics{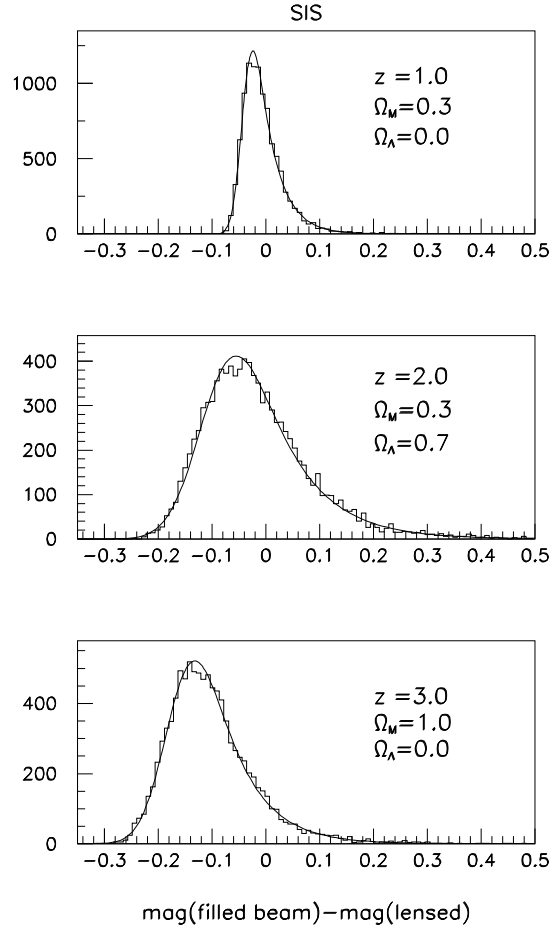}}
  \caption{Fits of magnification distributions for SIS lenses using
    the analytical expression in
    Eq.~(\ref{eq:gausexp2}).}\label{fig:sisdis} 
\end{figure}

\begin{figure}
\resizebox{\hsize}{!}{\includegraphics{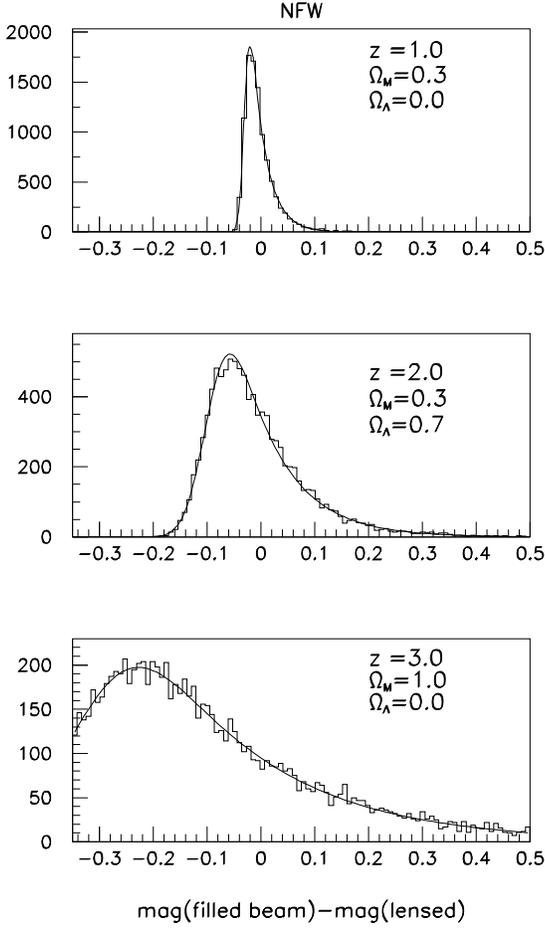}}
\caption{Fits of magnification distributions for SIS lenses using the
  analytical expression in Eq.~(\ref{eq:gausexp2})}\label{fig:nfwdis}
\end{figure}

In Fig.~\ref{fig:sisdis} the fits to the magnification distributions
are shown for SIS lenses for source locations $z=1$, $2$ and $3$ and
for three different cosmologies. In Fig.~\ref{fig:nfwdis} the
same source locations and cosmologies are used but the lenses
are of NFW type. 


\begin{figure}
\resizebox{\hsize}{!}{\includegraphics{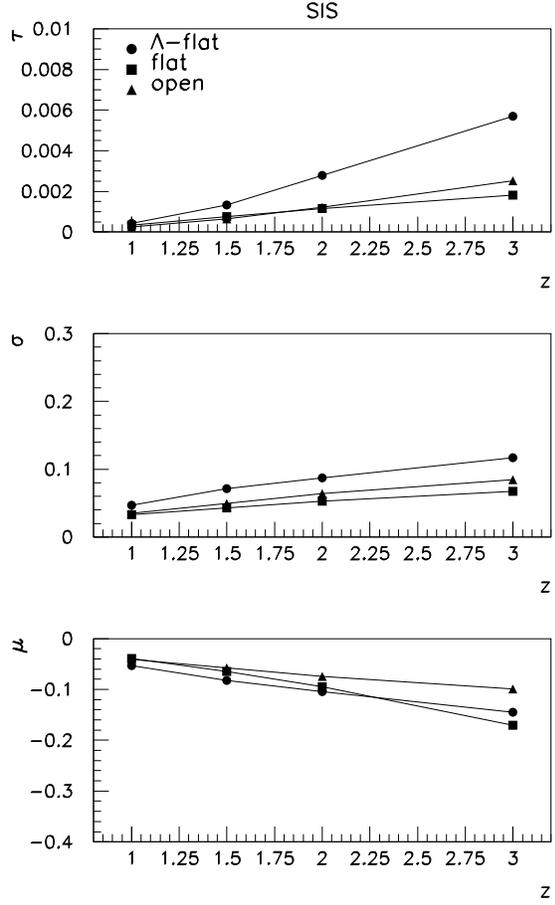}}
  \caption{Redshift and cosmology dependence of parameters in
    Eqs.~(\ref{eq:gausexp1}) and (\ref{eq:gausexp2}) for SIS
    lenses.}\label{fig:sispar} 
\end{figure}

\begin{figure}
  \resizebox{\hsize}{!}{\includegraphics{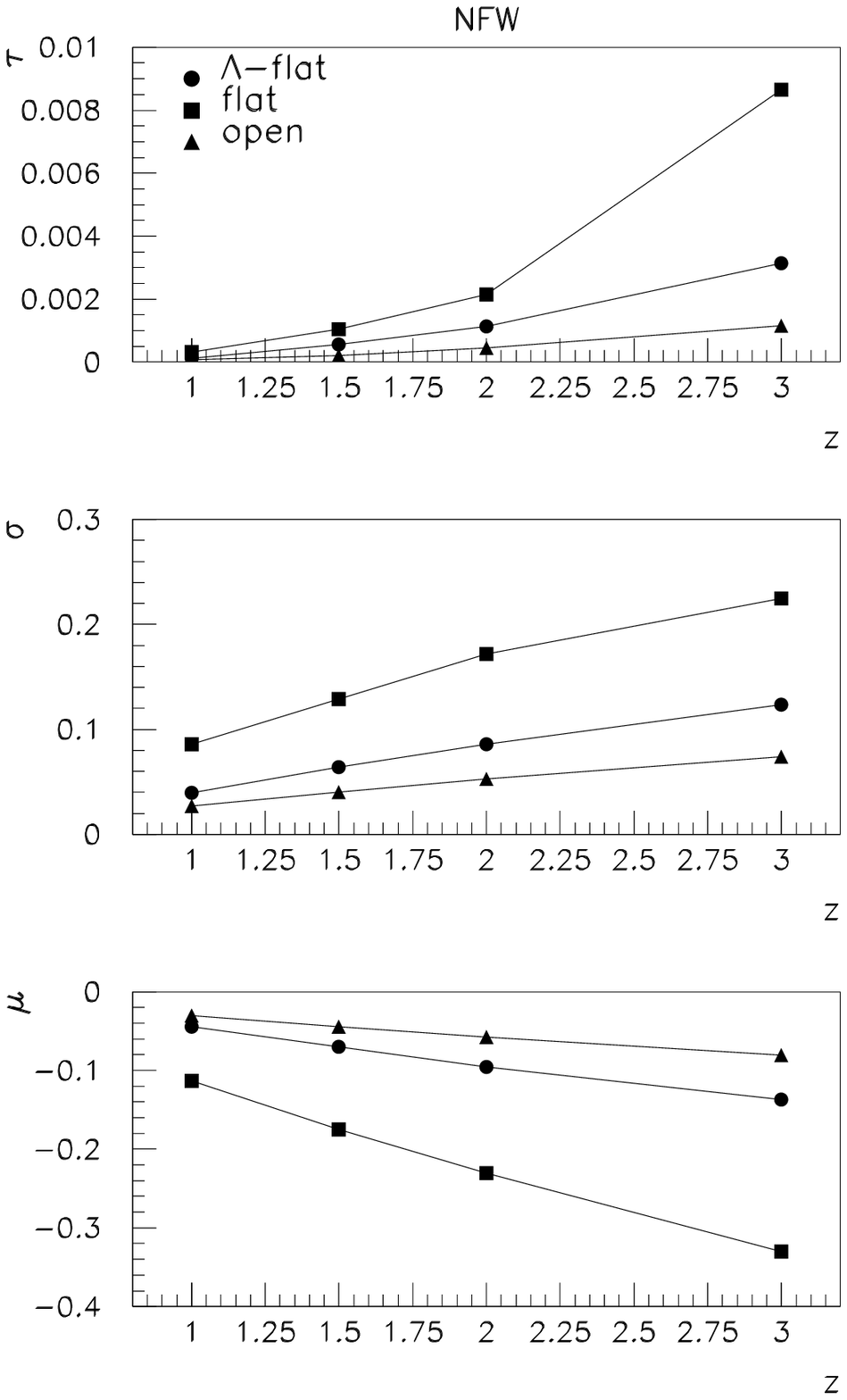}}
  \caption{Redshift and cosmology dependence of parameters in
    Eqs.~(\ref{eq:gausexp1}) and (\ref{eq:gausexp2}) for NFW
    lenses.}\label{fig:nfwpar} 
\end{figure}

The parametrization of Eqs.~(\ref{eq:gausexp1}) and (\ref{eq:gausexp2}) 
may be used to study the shape of the magnification distribution as a
function of redshift and cosmological parameters, as shown in 
Figs.~\ref{fig:sispar} and \ref{fig:nfwpar}.

The estimates of the cosmological parameters may be further refined
and the understanding of the mass distribution in galaxy halos could
be deduced by fitting  the parameters $\tau$, $\sigma$ and $\mu$ to
the residual magnitude distributions in future high-redshift supernova
searches, such as the SNAPsat project or at NGST. The accuracy of this
method will be discussed in a forthcoming paper
(Bergstr\"om et al., in preparation).

\section{Discussion}\label{sec:disc}

In this paper, the method of Holz \& Wald (\cite{art:HolzWald1998})
has been generalized to
include the effect of non-vanishing pressure. A motivation for this
exercise is to use this method as part of a large package  
 for a detailed simulation of most aspects 
high-redshift supernova determination of the geometry
of the universe (Bergstr\"om et al., in preparation). The possibility to allow
for various matter contributions is then desirable. Another improvement
has been to allow for a more realistic mass profile of the
galaxy halos, the Navarro-Frenk-White distribution, where we have
given analytical formulas for the geodesic deviation integrals.
With this improvement, and by adding matter in our model universes
according to a Schechter function with parameters fixed by
observation, we think we have achieved a sufficient degree of
sophistication to make realistic predictions of the lensing effects
of present and future deep supernova searches.

Our results show, in agreement with previous work 
(Holz \& Wald \cite{art:HolzWald1998}; Metcalf \& Silk \cite{metcalf})
that the (unlikely) case of having most of the dark matter distributed
in the form of point masses can be quite easily distinguished 
from more realistic halo models, if a sufficient number of supernovae
can be detected at large redshifts. It will be more difficult to
distinguish between a singular isothermal sphere model and the N-body
results
of Navarro, Frenk and White, unless $\Omega_M$ is large and one can
measure supernovae out to $z\sim 2$ (see Fig.~\ref{fig:model-7}).
However, this also means that our method of determining the properties
of the lens population by matching the measured Schechter function with
the required $\Omega_M$ gives quite robust results for the predicted
magnifications of Type Ia supernovae over a broad range of reasonable
halo models. This will be helpful when making realistic predictions
for future deep supernova searches.

\begin{acknowledgements}
The authors would like to thank Daniel Holz and Bob Wald for helpful
comments, and Julio Navarro for providing his numerical code
relating the parameters of the NFW model. 
\end{acknowledgements}

\appendix

\renewcommand\thesection{Appendix \Alph{section}}

\section{: $J^\alpha\!_\beta$ for the Navarro-Frenk-White matter
  distribution}\label{app:JNFW}
In this Appendix, we give analytic expressions for the integrals
(\ref{eq:jab})
needed for the evolution of the optical matrix for a bundle of 
light which passes near a mass distribution of the NFW form.
We assume a spherically symmetric matter distribution, the impact parameter 
is $b$, the characteristic radius of the NFW model is $R_s$ and
the cell radius is $R_c$.
When $R_s\neq b$, it follows that
\begin{eqnarray}
  (J^{\rm NFW})^X\!_X&=&2 C \left\{
  \frac{\sqrt{R_c^2-b^2}(R_s^2-2R_c^2-b^2)}{R_c^2(R_s+R_c)(R_s^2-b^2)}
  \right. \nonumber \\ && 
  +\frac{2}{b^2}
  \ln\left[\frac{R_c}{b}+\sqrt{\left(\frac{R_c}{b}\right)^2-1}\right]
  \nonumber \\ && 
  -\frac{\sqrt{R_c^2-b^2}(2R_c^2+b^2)}{b^2R_c^3}
  \ln\left(1+\frac{R_c}{R_s}\right) 
  \nonumber \\ && \left.
  +\frac{2R_s(R_s^2-2b^2)}{b^2|R_s^2-b^2|^{3/2}}f(b,R_s,R_c)\right\} ,
\end{eqnarray}
\begin{eqnarray}
(J^{\rm NFW})^Y\!_Y &=& -2 C \left\{
  \frac{\sqrt{R_c^2-b^2}}{R_c^2(R_s+R_c)} 
  \right. \nonumber \\ && 
  +\frac{2}{b^2}
  \ln\left[\frac{R_c}{b}+\sqrt{\left(\frac{R_c}{b}\right)^2-1}\right]
  \nonumber \\ &&
  -\frac{\sqrt{R_c^2-b^2}(2R_c^2+b^2)}{b^2R_c^3}
  \ln\left(1+\frac{R_c}{R_s}\right)
  \nonumber \\ && \left.
  +\frac{2R_s\,{\rm sgn}(R_s-b)}{b^2|R_s^2-b^2|^{1/2}}f(b,R_s,R_c)\right\} ,
\end{eqnarray}
\begin{eqnarray}
  (J^{\rm NFW})^X\!_Y &=&(J^{\rm NFW})^Y\!_X=0\nonumber\\
  f(b,R_s,R_c)&=&\left\{
  \begin{array}{l}
  \!\!\ln\!\left(\!\frac{b^2+R_sR_c-\sqrt{R_s^2-b^2}\sqrt{R_c^2-b^2}}
  {b(R_s+R_c)}\!\right) \, (R_s>b) \\
  \arctan\left(\frac{\sqrt{b^2-R_s^2}\sqrt{R_c^2-b^2}}
  {b^2+R_sR_c}\right) \quad (R_s<b)
  \end{array}\right. \nonumber ,
\end{eqnarray}
and the limit $R_s\rightarrow b$ is:
\begin{eqnarray}
  (J^{\rm NFW})^X\!_X &=&2 C \left\{
  \frac{2}{b^2}
  \ln\left[\frac{R_c}{b}+\sqrt{\left(\frac{R_c}{b}\right)^2-1}\right]
  \right.\nonumber\\
  &&-\frac{\sqrt{R_c^2-b^2}(2R_c^2+b^2)}{b^2R_c^3}
  \ln\left(1+\frac{R_c}{b}\right)-  \nonumber\\
  &&\left.\left(\frac{R_c-b}{R_c+b}\right)^{3/2}
  \frac{R_c^2+3(R_c+b)^2}{3R_c^2b^2} \right\} ,
  \nonumber\\
  (J^{\rm NFW})^Y\!_Y &=&-2 C \left\{
  \frac{\sqrt{R_c^2-b^2}}{R_c^2(R_c+b)}
  \right. \nonumber \\ &&
  +\frac{2}{b^2}
  \ln\left[\frac{R_c}{b}+\sqrt{\left(\frac{R_c}{b}\right)^2-1}\right]
  \nonumber \\ &&
  -\frac{\sqrt{R_c^2-b^2}(2R_c^2+b^2)}{b^2R_c^3}
  \ln\left(1+\frac{R_c}{b}\right)
  \nonumber \\ && \left. 
  -2\frac{\sqrt{R_c^2-b^2}}{b^2(R_c+b)} \right\} .
\end{eqnarray}
The constant $C$ is given by
\begin{equation}
  C=\rho_{\rm crit}\delta_cR_s^3=
  \frac{M}{\ln(1+x_c) - \frac{x_c}{1+x_c}} ,
\end{equation}
where $x_c=R_c/R_s$.

\section{: Parameter values for mass distribution}\label{app:parametervalues}

Here, we have compiled various parameter values, relevant for the mass
distribution discussed in Sect.~\ref{sec:massdist}. In what follows, absolute
magnitudes are denoted by ${\cal M}$, while masses are denoted by $M$.

\begin{description}
  \item[{\bf Lin et al.}] See Eq.~(32) of (Lin et al. \cite{lin}).
  \begin{itemize}
    \item[$\bullet$]  $\alpha=-0.70\pm 0.05$ 
    \item[$\bullet$]  ${\cal M}_*-5\log h=-20.29\pm 0.02$
    \item[$\bullet$]  $n_*=(1.9\pm 0.1)\cdot 10^{-2}\, h^{3}{\rm \ Mpc}^{-3}$
  \end{itemize}
  Lin et al. gives its range of validity as
  $-23.0\leq{\cal M}\leq-17.5$, which is about $\pm2.7$ 
  magnitudes around their characteristic magnitude $-20.29$. 
  This translates into a luminosity range $L\in L_*10^{0.4\cdot2.7}$, 
  and a mass range $M\in M_*10^{0.4\cdot2.7/(1-\beta)}$, which, with 
  $\beta=0.2$, gives $M/M_*\in [0.045,22.4]$. 

  \item[{\bf Kirshner et al.}] (Kirshner et al. \cite{KI83}; KI)
  \begin{itemize}
    \item[$\bullet$]  $\alpha_J= -1.25$ \quad (KI\,19)\\
    $\alpha_F=-1.25$ \quad (KI\,22)
    \item[$\bullet$]  ${\cal M}_{*,J}=-21.70$ \quad (KI\,18)\\
    ${\cal M}_{*,F}=-22.70$ \quad (KI\,21)
    \item[$\bullet$]  $L_{*}=1.0\cdot 10^{10}h^{-2}L_{\odot}$
    (see Binney \& Tremaine \cite{BT87}).
    \item[$\bullet$]  $n_{*}=1.2\cdot 10^{-2}h^{3}{\rm \ Mpc}^{-3}$ 
    \quad (KI\,23)
  \end{itemize} 

  \item[{\bf Peebles}] (Peebles \cite{book:Peebles}; PE)
  \begin{itemize}
    \item[$\bullet$]  $\beta =0.2$ \quad (PE\,3.39)
    \item[$\bullet$]  $\alpha =-1.07\pm 0.05$ \quad (PE\,5.130)
    \item[$\bullet$]  ${\cal M}_{*}=-19.53\pm 0.25+5\log h$ in the $b_{J}$
    magnitude system where ${\cal M}_{\odot}=5.48$. 
    \quad (PE\,5.139,\,5.140)
    \item[$\bullet$]  $L_{*}=1.0\cdot 10^{10}e^{\pm 0.23}h^{-2}L_{\odot}$
    \quad (PE\,5.141)
    \item[$\bullet$]  $n_{*}=1.0\cdot 10^{-2}e^{\pm 0.4}h^{3}{\rm \ Mpc}^{-3}$
    \quad (PE\,5.142)
    \item[$\bullet$]  $v_{*}=220\mbox{ km/s}$ \quad (PE\,3.35)
    \item[$\bullet$]  $\lambda =0.25$ \quad (PE\,3.35)
  \end{itemize} 
\end{description}
In the numerical work in this paper, we have used 
\begin{itemize}
\item[$\bullet$] $\beta =0.2$
\item[$\bullet$] $\alpha =-0.7$
\item[$\bullet$] $L_{{\rm min}}=0.5 L_{*}$
\item[$\bullet$] $L_{{\rm max}}=2.0 L_{*}$
\item[$\bullet$] $n_*=1.9\cdot 10^{-2}\, h^{3}{\rm \ Mpc}^{-3}$
\item[$\bullet$] $v_{*}=220\mbox{ km/s}$ 
\item[$\bullet$] $\lambda =0.25$
\end{itemize}

\section{: Mean area vs. luminosity}\label{app:meanarea}

Holz and Wald (\cite{art:HolzWald1998}) use the fact that the area of
the boundary of the past of 
an observer should be very nearly equal to the area of the past light 
cone in the underlying FL model. From this they deduce that adding up 
all areas -- counting beams with an even number of caustics as positive
and beams with an odd number as negative -- one should very nearly 
obtain the FL result, i.e., $\langle A_{0}\rangle \simeq A_{\rm FL}$,  
where the zero-subscript denote unweighted quantities. Here we 
investigate whether this claim is
consistent with the fact that the average magnification should be 
equal to the FL magnification. 

Imagine that, without weighting, we 
get an area distribution $n_{0}(A)$, where we assume that lensing 
effects are weak enough to only give primary images. 
According to HW, we then have
\begin{equation}
    \langle A_{0}\rangle =\frac{\int n_{0}(A)AdA}{\int 
    n_{0}AdA}=A_{\rm FL}.
    \label{eq:AHW}
\end{equation}
In our simulations, the area distribution will be a convolution of 
$n_{0}(A)$ and a probability distribution $p\,(A)$,
\begin{equation}
    n(A)=n_{0}(A)p\,(A)=n_{0}(A)\frac{A}{A_{\rm empty}}.
    \label{eq:n}
\end{equation}
Since we want the harmonic mean of areas to be equal to the FL value, 
we get
\begin{eqnarray}
    \frac{1}{\langle A\rangle}&=&\frac{\int^{A_{\rm empty}}n(A)/A\,dA}
    {\int^{A_{\rm empty}}n(A)\,dA}\nonumber\\ 
    &=&\frac{\int^{A_{\rm
    empty}}n_{0}(A)/A_{\rm empty} dA}
    {\int^{A_{\rm empty}}n_{0}(A)A/A_{\rm empty}dA}=\frac{1}{\langle
    A_{0}\rangle}.
    \label{eq:AA}
\end{eqnarray}
That is, the two consistency checks at least show consistency in the 
case where we only have primary images.  
The inclusion of non-primary images is highly non-trivial
since the concept of negative areas is difficult to interpret 
as luminosities. One possibility is to combine 
this approach with the one-lens approximation method for multiple 
image correcting. 
However, this is difficult to do analytically since one would then need 
the form of $n_{0}(A)$.

\end{document}